\documentclass[useAMS,usenatbib,usegraphicx]{mn2e}

\usepackage[utf8]{inputenc}
\usepackage[T1]{fontenc}
\usepackage{amsmath}
\usepackage{amssymb}
\usepackage{mathptmx} 

\usepackage[usenames,dvipsnames]{color}
\newif\ifusehyperref
\usehyperreftrue 
\ifusehyperref
 \usepackage[unicode]{hyperref}
 \usepackage[all]{hypcap}
 \hypersetup{backref,
             bookmarks,
             dvips,
             pdftitle=The Stellar Kinematic Signature of Massive Black Hole Binaries,
             pdfsubject=Astrophysics \& Cosmology,
             pdfkeywords={black hole physics -- galaxies : nuclei -- stellar dynamics},
             pdfdisplaydoctitle,
             colorlinks=true,
             citecolor=Blue,
             pdfauthor=Yohai Meiron,
             unicode}
 \let\OLDS\S
 \renewcommand{\S}{\textcolor{red}{\OLDS}} 
\else
 \newcommand{\href}[2]{#2}
\fi

\title[The Stellar Kinematic Signature of Massive Black Hole Binaries]{The Stellar Kinematic Signature of Massive Black Hole Binaries}
\author[Yohai Meiron and Ari Laor]{Yohai Meiron\thanks{E-mail: \href{mailto:ym@physics.technion.ac.il}{ym@physics.technion.ac.il} (YM); \newline\href{mailto:laor@physics.technion.ac.il}{laor@physics.technion.ac.il} (AL)}
and Ari Laor\footnotemark[1]\\
Department of Physics, Technion -- Israel Institute of Technology, Haifa 32000, Israel}
\begin{document}

\date{Accepted 2010 May 16. Received 2010 May 03; in original form 2010 February 11}

\pagerange{\pageref{firstpage}--\pageref{lastpage}} \pubyear{2010}

\maketitle

\label{firstpage}

\begin{abstract}
The stalling radius of a merging massive binary black hole (BBH) is expected to be below 0\farcs1 even in nearby galaxies \citep{yu02}, and thus BBHs are not expected to be spatially resolved in the near future. However, as we show below, a BBH may be detectable through the significantly anisotropic stellar velocity distribution it produces on scales 5--10 times larger than the binary separation. We calculate the velocity distribution of stable orbits near a BBH by solving the restricted three body problem for a BBH embedded in a bulge potential. We present high resolution maps of the projected velocity distribution moments, based on snapshots of $\sim 10^8$ stable orbits. The kinematic signature of a BBH in the average velocity maps is a counter rotating torus of stars outside the BBH Hill spheres. The velocity dispersion maps reveal a dip in the inner region, and an excess of 20--40\% further out, compared to a single BH of the same total mass. More pronounced signatures are seen in the third and fourth Gauss--Hermite velocity moments maps. The detection of these signatures may indicate the presence of a BBH currently, or at some earlier time, which depends on the rate of velocity phase space mixing following the BBH merger.
\end{abstract}


\section{Introduction}

The discovery that most galaxies harbour a massive black hole (BH) at their core \citep{1998AJ....115.2285M}, and the commonly accepted interpretation of cosmological structure formation simulations, that galaxies grow by mergers (e.g. \citealt{1993MNRAS.264..201K}, cf. \citealt{2006MNRAS.368....2D} that growth is mostly by gas accretion) implies that massive binary black holes should commonly form at the core of galaxies \citep{2003ApJ...582..559V}. Unlike the previous view, that the formation of such a binary results from the rare process of merging during the life of the rare objects, quasars \citep*{b80}. Dynamical friction on the background stars leads to inspiral of the binary on a dynamical time-scale, until the binary becomes hard, which occurs at a binary separation close to
\begin{equation}
a_{\rm h}\equiv\frac{q}{1+q}\frac{GM_\bullet}{4\sigma^2},\label{eq:ah}
\end{equation}
where $G$ is the gravitational constant, $M_\bullet$ is the mass of the primary BH, $q$ is the mass ratio of the BHs ($q\leq1$), and $\sigma$ is the bulge 1D velocity dispersion. The inspiral rate slows drastically when $a \lesssim a_{\rm h}$ (e.g. \citealt{2006ApJ...648..976M}), and further inspiral is set by the rate at which stars diffuse in phase space into the loss cone \citep{1976MNRAS.176..633F}\footnote{More properly termed the loss cylinder, as pointed out by \citet{1978ApJ...226.1087C}. See also \S\ref{sec:stability-maps} below.}. The minimum phase space diffusion rate is set by the two body relaxation rate. This mechanism may be fast enough in the lowest luminosity bulges, where the stellar cores are the densest, and may lead to merger through gravitational wave emission in less than the Hubble time (but see recent suggestions that the core density in the Milky Way is smaller than originally thought, \citealt{2010arXiv1001.5435M} and references therein) . In more luminous bulges, the binary inspiral is expected to stall (e.g. \citealt{2004ApJ...602...93M}; \citealt{2006RPPh...69.2513M}, and references therein), leading to the  final parsec problem \citep{b80}. The largest plausible stalling radius in the nearest galaxies is just below $0\farcs1$ \citep{yu02}, and is thus generally unresolved. The stalling may be overcome by diffusion of orbits into the loss cone induced by tangential forces. Either due to bulge triaxiality or bar like structure, or by massive perturbers in the form of molecular clouds or a third BH due to another merger (e.g. \citealt{2004ApJ...606..788M}; \citealt{2006ApJ...642L..21B}; \citealt{2007MNRAS.377..957H}; \citealt{tal07}; \citealt{2008ApJ...677..146P}). Alternatively, inspiral may be induced by angular momentum loss of the BBH to circumbinary gas (e.g. \citealt{2005ApJ...634..921A}; \citealt{2007Sci...316.1874M}; \citealt{2009MNRAS.393.1423C}). These processes may lead to a merger on time-scales well below the Hubble time.

The formation process of the BBH ejects stars from the bulge. This occurs on scales significantly larger than $a_{\rm h}$, and it flattens the stellar density profile. This BBH ``scouring'' may be responsible for the formation of the core present in luminous ellipticals (\citealt{mm01}; \citealt{mm05}). Recent high quality observations by \citet{2009ApJ...691L.142K} indicate a remarkably tight correlation between the deduced mass deficiency in the core region, and the black hole mass. This correlation is likely a signature of the cumulative scouring effects of merging BHs during the BH growth. Are there any other signatures induced by the merger?

If the binary stalls, then a significant fraction of galaxies may harbour a compact BBH (e.g. \citealt{2003ApJ...582..559V}), which cannot currently be directly resolved \citep{yu02}. The binary affects the stability of stellar orbits on scales up to several times larger than the binary separation. The presence of a binary may thus be inferred through its signature on the background stellar distribution and kinematics (e.g. \citealt{2005CQGra..22S.347C}; Kandrup et al. 2003). The purpose of this paper is to determine this signature and provide a tool which can be used to detect the effect of a binary black hole on scales 5--10 times larger than $a_{\rm h}$, a scale which may be resolved in nearby galaxies with current angular resolutions.

We first explore the stability of orbits near a BBH by calculating stability maps. These maps present the time it takes for a test particle to become unbound, as a function of its position in velocity space, at a given launching radius. A similar concept appears in \citet{1997AJ....113.1445W} in the context of small bodies near a binary stellar system, but the coordinates used there were the semi-major axis and the inclination of the orbit, which are less relevant for the kinematic signature explored here. The stability maps elucidate how the velocity phase space populated by stable orbits evolves with integration time, and how it varies with distance from the binary. The maps shows how the velocity distribution function $f(v)$ evolves from a smooth isotropic function at a distance $r\gg a_{\rm h}$, to a highly anisotropic function, which produces the BBH kinematic signature.

We derive the BBH kinematic signature by integrating orbits of test particles around a massive BBH, i.e. by solving the restricted 3-body problem, for a binary in a circular orbit, embedded in a bulge potential. We take snapshots of the system once it approaches a steady state, and produce maps of the projected distribution of the stars and their velocity distributions. Similar scattering experiments were already done in the past (e.g. \citealt{hills83}; \citealt{mikola92}; \citealt{1996NewA....1...35Q}; \citealt{2007ApJ...660..546S}). The purpose of these studies was to derive the effect of the stellar background on the BBH merger rate, while our purpose is the reverse, to derive the effects of the BBH on the background stellar population. The latter calculations were already carried out by \citet{mm01} based on $N$-body simulations. These simulations were limited to $N\lesssim10^5$ and cover a region more than $10^3$ larger than $a_{\rm h}$, thus they do not allow accurate mapping of the projected line of sight velocity distributions (LOSVDs) on scales of a few times $a_{\rm h}$, as done here. Since the background stellar mass within $a_{\rm h}$ is negligible (see below), and 2-body scattering cross sections are still small, $N$-body simulations can be replaced by the much faster scattering experiments. These experiments allow us to probe the stellar kinematics with a roughly $10^5$ higher resolution, in the relevant region, compared to \citet{mm01}. In \S\ref{sec:methods} we present the method of the calculation, and in \S\ref{sec:results} we present results on the phase space stability regions, maps of the projected velocity distribution moments, velocity moments along some slit positions, various LOSVDs, and also stellar density profiles and projected density maps. The results are discussed in \S\ref{sec:discussion}, and summarized in \S\ref{sec:conclusions}.

\section{Model}

\subsection{Orbital Elements}

The two BHs are assumed to be in circular orbits in the $x$-$y$ plane, about their centre of mass at $(x,y)=(0,0)$. The orbital period, the relative velocities, and radii are:
\begin{eqnarray}
T & = & 2\pi \sqrt{\frac{a^3}{GM_\bullet (1+q)}} \label{eq:T}\\
V & = & \sqrt{(1+q)\frac{GM_\bullet}{a}}\label{eq:V}\\
R_1 & = & \left(\frac{q}{1+q}\right)a\\
R_2 & = & \left(\frac{1}{1+q}\right)a \label{eq:R2}
\end{eqnarray}
where $a$ is the binary separation, corresponding to the stalling radius. We take $a=a_{\rm h}$, i.e. the stalling separation is the hard binary separation (defined in equation \ref{eq:ah}). The recent simulations of \citet{2006ApJ...648..976M} (table 1 there) indicate that the stalling radius agrees with $a_{\rm h}$ to typically 20\%. Combining equations (\ref{eq:ah}) and (\ref{eq:V}) gives:
\begin{equation}
\frac{V}{\sigma}=\frac{2(1+q)}{\sqrt{q}}\label{eq:v0-sigma}
\end{equation}

The physical scales are set by specifying two parameters, say $M_\bullet$ and $V$. However, since  $M_\bullet$ and $\sigma$ are correlated through the $M$-$\sigma$ relation (\citealt{2000ApJ...539L..13G}; \citealt{2000ApJ...539L...9F}; \citealt{2002ApJ...574..740T}, and citations thereafter), only one parameter is required. We use the recent \citet{2009ApJ...698..198G} relation:
\begin{equation}
\log M_8 =0.12\pm 0.08 + (4.24\pm 0.41)\log\sigma_{200}
\end{equation}
where $M_8=M_\bullet/10^8~{\rm M}_\odot$ and $\sigma_{200}=\sigma/200~{\rm km~s}^{-1}$, which gives:
\begin{eqnarray}
a_{\rm h} & = & (3.1\pm 0.3)\frac{q}{1+q}M_8^{0.53\pm 0.05}~{\rm pc}\label{eq:ah-m}\\
    & = & (3.5\pm 0.3)\frac{q}{1+q}\sigma_{200}^{2.24\pm 0.41}~{\rm pc}.
\end{eqnarray}
Thus, the binary separation is expected to be on the parsec scale. Similarly, the orbital period is:
\begin{eqnarray}
T & = & (5.5\pm 0.3)\times 10^4\frac{q^{1.5}}{(1+q)^2}M_8^{0.29\pm 0.03}~{\rm yr}\\
    & = & (5.0\pm 0.3)\times 10^4\frac{q^{1.5}}{(1+q)^2}\sigma_{200}^{1.24\pm 0.1}~{\rm yr}.
\end{eqnarray}
Thus, a characteristic orbital time-scale of $10^4$ years, with a relatively weak dependence on mass.

\subsection{Model Units}\label{sec:units}

For the sake of simplicity, we use the following values $G=1$, $M_\bullet=1$, and $a=2$. To convert back to physical units, one needs to specify the physical value of $M_\bullet$; and then the length, which is measured in units of $a_{\rm h}/2$, is given in parsecs by equation (\ref{eq:ah-m}). Time is then measured in units of $(a_{\rm h}^3/8GM_\bullet)^{1/2}$ and velocity in units of $(2GM_\bullet/a_{\rm h})^{1/2}$. Alternatively, the conversion from dimensionless to physical units can also be made using the value of $\sigma_{200}$.

Note that models with different $q$ correspond to different physical scales for the same mass scaling (see equation \ref{eq:ah}). Table \ref{tab:units} gives the physical scales of the model units for $M_\bullet=10^8~{\rm M}_\odot$, and $q=1$ and 0.1 used below. Also note that $\sigma$ in model units is a function of $q$ only (equation \ref{eq:v0-sigma}).

\begin{table}
\caption{The orbital elements in model units and physical units. The conversion is for $M_\bullet = 10^8~{\rm M}_\odot$, and the conversion factors are listed in the lower part of the table.}
\label{tab:units}
\begin{tabular}{cccccc}
\hline
 & & \multicolumn{2}{c}{$q=1$} & \multicolumn{2}{c}{$q=0.1$}\\
 & Expression & Model & Physical & Model & Physical \\
\hline
$a$ & 2 & 2 & 1.55~pc & 2 & 0.28~pc \tabularnewline
$T$ & $4\pi\sqrt{\frac{2}{1+q}}$ & 12.57 & 12\,783~yr & 16.94 & 1\,336~yr \tabularnewline
$V$ & $\sqrt{\frac{1+q}{2}}$ & 1 & 745~km/s & 0.74 & 1\,296~km/s \tabularnewline
$\sigma$ & $\sqrt{\frac{q}{8(1+q)}}$ & 0.25 & 186~km/s & 0.11 & 186~km/s \tabularnewline
\hline
\hline
 & Length & 1 & 0.78~pc & 1 & 0.14~pc \\
 & Time & 1 & 1017~yr & 1 & 79~yr \\
 & Velocity & 1 & 745~km/s & 1 & 1\,747~km/s \\
\hline
\end{tabular}
\end{table}

\subsection{Bulge Properties}\label{sec:bulge}

The BBH is embedded in an isothermal sphere, i.e. the velocity distribution of the stars is a Maxwell--Boltzmann distribution $f(v)=Av^2\exp^{-v^2/2\sigma^2}$, where $v$ is the magnitude of the velocity, $A$ is a normalization coefficient. The assumed stellar density profile is $\rho(r)=\sigma^2/2\pi G r^2$, the self-consistent solution for a singular isothermal sphere. This is only an approximate solution for a finite mass bulge, and is not appropriate within the BH sphere of influence. Here we use $f(v)$ and $\rho(r)$ as convenient initial conditions for the stellar distribution close to the BBH. Also, $\rho(r)$ allows a convenient representation of the bulge potential, and is surprisingly accurate for the average properties of massive ($>3\times 10^{10}~{\rm L}_\odot$) ellipticals, within their effective radius \citep{2009ApJ...703L..51K}. To avoid the non physical divergence of $\rho(r)$ at $r=0$, we assume a core structure:
\begin{equation}
\rho(r)=
\begin{cases}
 \rho_0 & r<h\\
 \rho_0\left(\frac{h}{r}\right)^{2} & r>h
\end{cases}\label{eq:isothermal-density}
\end{equation}
where $h$ is an arbitrary break radius and
\[
\rho_0=\frac{\sigma^2}{2\pi G h^2}.
\]
The integrated bulge mass derived from the above density field is:
\begin{equation}
M(r)=
\begin{cases}
 \frac{2 \sigma^2}{3Gh^2}r^3 & r<h\\
 \frac{2 \sigma^2}{3G}(3r-2h) & r>h
\end{cases}\label{eq:isothermal-mass}
\end{equation}
The expression for the gravitational potential (i.e. the bulge potential) is:
\begin{equation}
\Phi_{\rm bulge}(r)=
\begin{cases}
 \frac{\sigma^2}{3Gh^2}r^2 & r<h\\
 \frac{\sigma^{2}}{G}\left[\frac{4h}{3r}+2\ln\left(\frac{r}{h}\right)-1\right] & r>h
\end{cases}\label{eq:bulge-potential}
\end{equation}

The $N$-body merger simulations of \citet{mm01} indicate that at the time the binary becomes hard, for a short while, an $r^{-2}$ density profile extends down to the scale of $a_{\rm h}$\footnote{Though this is not necessarily a realistic result, given the small number of $10^3$ stars interacting with the BH in their simulation within $r_{\rm infl}$, and the implied very short relaxation time.}. We therefore assume $h=a/2$ (in model units, $h=1$, but note the final larger core radius produced at the end of the simulation). From here on, we work only in model units (defined in \S\ref{sec:units}).

The calculations of the binary orbital elements (equations \ref{eq:T}--\ref{eq:R2}) ignores the bulge mass within the BBH orbits. The enclosed mass within the radius of the secondary BH (equation \ref{eq:R2}), which resides outside the uniform density core, is:
\begin{equation}
M(R_2)=\frac{q(2-q)}{6 (1+q)^2}.
\end{equation}
Thus, $M(R_2)\lesssim 0.04M_\bullet$, and the bulge mass was therefore neglected in the derivation of $V$ and $T$ made above.

The bulge mass grows linearly with $r$, and becomes larger than the combined BH mass at:
\begin{equation}
r_{\rm infl}=\frac{26}{3} + \frac{4}{q} + 4q.
\end{equation}
This definition of the radius of influence differs somewhat from the commonly used definition of $GM_\bullet/\sigma^2$, or $8/q+8$ in the model units. For $q=1$, $r_{\rm infl} \approx 17$ and for $q=0.1$,  $r_{\rm infl} \approx 49$; but note that in physical units the latter number is smaller (see Table \ref{tab:units}). The apparent divergence of $r_{\rm infl}$ for $q\rightarrow 0$ results from the divergence of $V$ as $a_{\rm h}\rightarrow 0$ (equations \ref{eq:ah}, \ref{eq:V}). Below we also simulate orbits around a single BH, for the purpose of comparison. This simulation is designated as the $q=0$ case, but we do not use the $V$ based normalization, due to its divergence.

\section{Methods}\label{sec:methods}

\subsection{Orbit Integration}\label{sec:integration}

The orbit of each test particle (representing a star) is solved using a 5$^{\rm th}$ order Runge--Kutta method with adaptive step size control. The fractional error tolerance in the code was set to $10^{-6}$, further lowering this value had no detectable effect on the results. The accuracy of the calculation was also verified through conservation tests of the value of the Jacobi integral, a constant of motion in the circular restricted 3-body problem.

The integration for each particle was terminated if it reached $r_\infty=200$. Integrating to higher values of $r_\infty$, up to $\sim 3\times 10^3$, yielded negligible effects on the results presented below. An upper limit on the physical distance of the order of $< 10^3~{\rm pc}$ is also expected as tangential forces produced by various deviations from the pure spherical symmetry, assumed here, become more likely and more effective in changing the particle's angular momentum. when moving far away from the centre.

The integration was also terminated if the particle reached $r_{\rm tidal}=10^{-3}$ from either BHs. This represents the tidal disruption of the star by the BH (the orbit is than termed as ``crashed''). The true tidal disruption radius is $r_*(2M_{\rm BH}/m_*)^{1/3}$, where $r_*$ is the radius of the star, which is about two orders of magnitude smaller than assumed here. However, setting $r_{\rm tidal}=10^{-3}$ was enough to set a negligible rate of stellar depletion compared to the rate of escape from the system (reaching $r_\infty$), and thus led to a negligible effects on the results. Reducing further $r_{\rm tidal}$ was also expensive in computing time, and was therefore avoided.

The integration was usually stopped at $t=10^4$, which corresponds to $\sim 800$ revolutions for a $q=1$ binary, or a physical time-scale of $\sim 10^7$ years. Orbits that have neither diverged (escaped) nor crashed before the end of their integration were defined as ``stable''. We also investigated the effects of extending the orbital stability to $t=10^6$, i.e. extending to $10^9$ years. As we show below, most unstable orbits diverge on a few orbital time-scales, and a steeply decreasing fraction of the phase space volume is populated by unstable orbits with increasing time-scale.

\subsection{Simulations}\label{sec:simulations}

To better understand the effects of a BBH on the LOSVDs, we first explore the evolution with time of the regions in velocity phase space populated by stable orbits, as a function of distance from the BBH. For this purpose we use {\it stability maps} (see \S\ref{sec:stability-maps}). In these maps, a fixed spatial position is chosen as the particle's starting point, and a 2D grid of initial velocity vectors in a certain plane is created. Each grid point is integrated up to $t=10^4$, or $t=10^6$, and if the integration is terminated earlier, the grid point is tagged according to the time it took for the orbit to terminate (reach $R_{\infty}$ or $r_{\rm tidal}$). The stability maps are also useful for understanding the effects of various parameters (e.g. the bulge potential) on the orbit stability, and are also a tool for the code development (locate inaccuracies and bugs throughout parameter space). 

The stability maps do not produce directly observable results. For this purpose we performed 3D Monte Carlo (MC) simulations to derive the observable kinematic signature. The initial positions of the particles in these simulations were drawn randomly from a $\rho(r)\propto r^{-2}$ distribution up to $r_{\rm max}=60$. It was verified that particles which were initiated at $r_{\rm max}>60$ (using a shell $60<r<80$) made a negligible change in the results. The decreasing effect of outer regions is expected since the line of sight integration of the $r^{-2}$ stellar distribution scales as $r^{-1}$. For the sake of simplicity we did not introduce a flat core at $r<h=1$ to the initial MC $r^{-2}$ distribution, as the $r<1$ region is only relevant for the spatially unresolved stellar population within the Hill sphere of each BH.

In velocity space, initial distributions in each direction were drawn from a normal distribution with variance $\sigma^2$, or equivalently a Maxwell--Boltzmann $f(v)$, as noted in \S\ref{sec:bulge}. A position dependent cutoff was applied so that $|v|<v_{\rm esc}$, where $v_{\rm esc}$ is the velocity required to reach $r_\infty$ from a given point in space (therefore, the initial $f(v)$ varies slightly with position). This cut is only for the sake of convenience, as all orbits with $|v|>v_{\rm esc}$ are found to be unstable (see the stability maps in \S\ref{sec:stability-maps}).

Orbits residing within the Hill spheres of the two BHs, with a total energy below the minimal potential energy at the Lagrangian points (as measured in the corotating system), cannot escape. All these orbits are therefore bound, but stars may be destroyed through tidal disruption (see \citealt{2009ApJ...697L.149C}), which is beyond the scope of this study (see \S\ref{sec:integration}). Because of the high accelerations in this region, the integration time of a single orbit can exceed by a few orders of magnitudes the integration time for an orbit outside the Hill spheres. We therefore separated each simulation into two parts: one for initial conditions within $r<2$ which includes almost all the Hill sphere orbits, and the other for $2<r<60$, which includes almost none. The two datasets were pieced together with the appropriate weights for an  $r^{-2}$ distribution. 

\section{Results}\label{sec:results}

\subsection{Stability Maps}\label{sec:stability-maps}

\begin{figure}
\begin{center}
\includegraphics[width=\columnwidth]{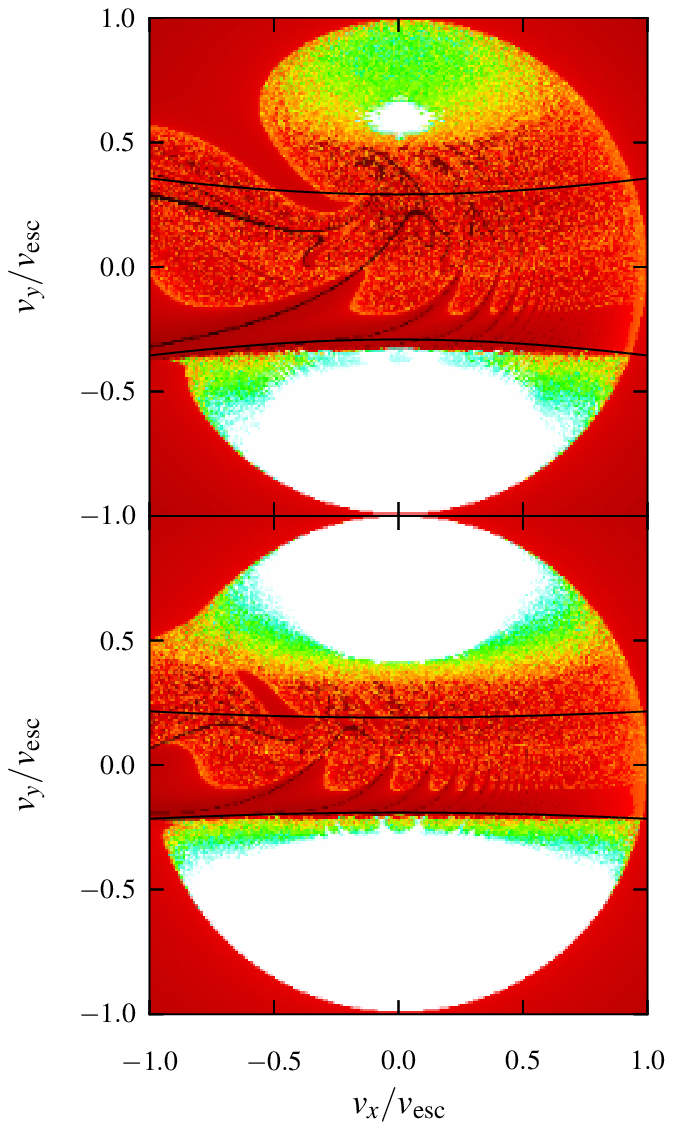}
\includegraphics{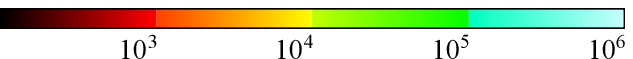}
\end{center}
\caption{The evolution of the stability maps with time, up to $t=10^6$, for a $q=1$ BBH. The horizontal and vertical axes correspond to the initial $v_x$ and $v_y$ velocities, normalized by the local $v_{\rm esc}$ (where $v_{\rm esc}=1.31, 1.06$ for $x=5, 10$). The orbits are launched from a point along the $x$-axis: $x=5$ (top panel) and $x=10$ (bottom), with $v_z=0$ (planar orbits). $v_y$ is therefore the tangential velocity, and positive values of which belong to corotating orbits. The colour indicates the survival time of each orbit. White represents stable orbits (survive to $t\geq 10^6$). The black lines bound the orbits within the classical loss cone. Note the near convergence of the stable areas for $t>10^5$. Below we integrate to $t=10^4$, which corresponds to a $\sim 10\%$ overestimate of the area in phase space occupied by stable orbits.}
\label{fig:stability-maps-longterm}
\end{figure}

Figure \ref{fig:stability-maps-longterm} presents cuts in phase space for $q=1$. Each point (or pixel) represents the initial velocity conditions of an orbit. A white pixel represents an orbit that neither diverged nor crashed throughout its integration; the orbits were followed to $t=10^6$. Non-white pixel is an orbit that became unstable at a time indicated by the colourbar. The upper and lower panels represent particles launched from $x=5,10$, respectively; in all panels $(y,z)=(0,0)$. The cuts in velocity space are in the $v_x$-$v_y$ plane, for orbits with $v_z=0$ (i.e. purely planar orbits). A total number of $256^2$ orbits were calculated in each map, where in each axis grid points are uniformly spaced in the range $-v_{\rm esc}<v_i<v_{\rm esc}$. Note that $v_{\rm esc}$ differs depending on the launching radius (see caption).

All particles with $v^2=v_x^2+v_y^2+v_z^2>v_{\rm esc}^2$ are unstable, as indicated by the circular boundary of the stable region, which has a radius of unity (Fig. \ref{fig:stability-maps-longterm}). Note that some orbits with $v>v_{\rm esc}$, which start inwards ($v_x<0$), do not escape immediately, in contrast to the outgoing orbits ($v_x>0$). The orbits which start inwards with $v>v_{\rm esc}$ can be temporarily trapped by the BBH, and wonder around on chaotic orbits. However, chaotic orbits are inherently unstable, and they inevitably lead to escape or a tidal disruption on time-scales of $10^2$-$10^3$ for these orbits.

The solid black lines correspond to the initial conditions required to reach a pericentre distance\footnote{We use the term pericentre distance to describe the radius of the closest point of a star from the centre of the coordinate system.}, $r_{\rm peri}$, of $r_{\rm min}=1$, assuming a purely central force (i.e. the two BHs are taken to be a single point source; the bulge force is unchanged). The line shape is given by:
\begin{equation}
v_y=\pm\sqrt{\frac{2\left[\Phi(x)-\Phi(r_{\rm min})\right]+v_x^2}{\left(x/r_{\rm min}\right)^2-1}},
\end{equation}
where $\Phi(r)=\Phi_{\rm bulge}(r)-(1+q)/r$ is the total potential, and the bulge term is given by equation (\ref{eq:bulge-potential}).

The orbits arrive to $\sim r_{\rm min}$, interact strongly with one of the two BHs, and are flung out. For negative values of $v_y$, corresponding to retrograde orbits (i.e. opposite to the BHs' rotation), the black lines fit well to the boundary of the stability region at $t=10^4$ (especially for the larger $r$, where the tangential part of the force is small). A similar boundary is seen at positive $v_y$, but corresponds to higher angular momentum than expected from the simplified loss cone solution.

As the integration time increases, from $t=10^3$ to $10^6$, the boundary of the stable regions in  Fig. \ref{fig:stability-maps-longterm} contracts. At $x=10$ there is a roughly uniform reduction in the stable area with each ten fold increase in time, indicating the reduction in area drops logarithmically with time. At $x=5$ there is a similar trend in the area of the stable retrograde orbits, but the prograde orbits area is roughly stable beyond $t=10^5$. The simulations described below were carried out to $t=10^4$, as the results get reasonably close to convergence on this time-scale. The asymmetry, which produces the BBH kinematic signature discussed in this paper, increases somewhat on time-scales longer than $10^4$. If the system is allowed to evolve further, the kinematic signature will be somewhat enhanced with respect to the results for $t=10^4$, presented below.

\begin{figure*}
\begin{center}
\includegraphics[width=\textwidth]{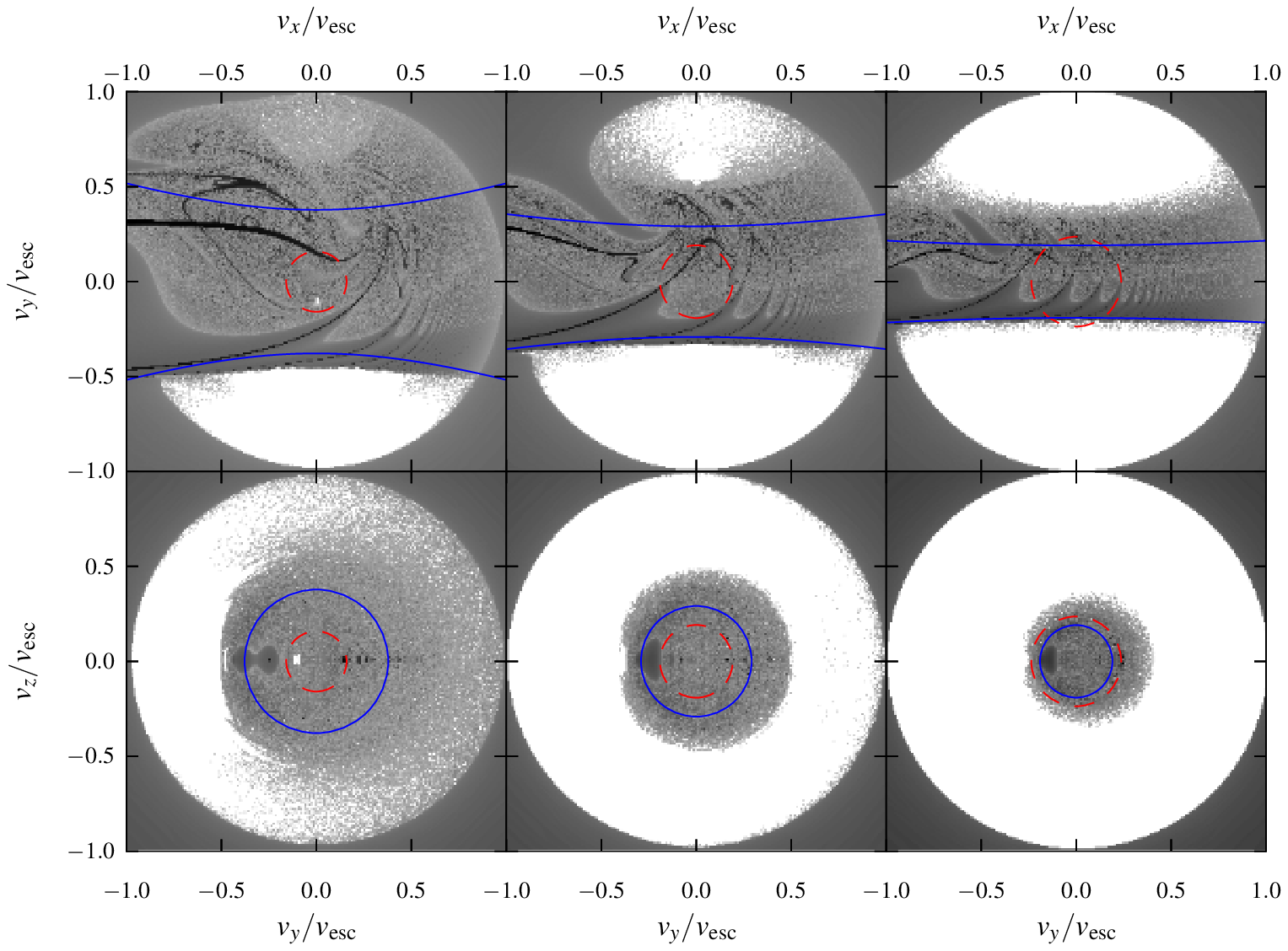}
\includegraphics{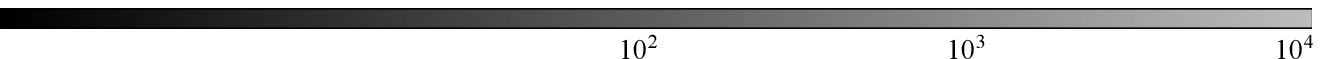}
\end{center}
\caption{Stability maps for cuts along different planes, as a function of distance from the centre. In both rows, the orbits are started on the $x$-axis and are launched from $x=3$ (left column), $x=5$ (middle) and $x=10$ (right). The integration time is $t=10^4$, for a $q=1$ BBH. The upper panels represent orbits in the $x$-$y$ plane ($v_z=0$), and the lower panels orbits start in the $y$-$z$ plane ($v_x=0$), i.e. start as purely tangential. Greyscale indicates the survival time of each orbit, and the white areas are stable orbits The panels show the initial velocities normalized by $v_{\rm esc}$ ($=1.57, 1.31, 1.06$ for $x=3, 5, 10$). Note the decreasing area of stable prograde orbits ($v_y>0$) with decreasing distance from the binary. Almost only retrograde orbits are present at $x=3$. The blue lines represent the border of the classical loss cone. It reproduces fairly well the boundary of the stable retrograde orbits, in particular in the $x$-$y$ plane, but fails significantly for the prograde orbits, already at $x=10$. The dashed red line represent the fixed $\sigma$ of the isothermal Maxwell--Boltzmann $f(v)$ used below.}
\label{fig:stability-maps-main}
\end{figure*}

In Figure \ref{fig:stability-maps-main} more phase space cuts are presented, for orbits followed to $t = 10^4$. The greyscale indicates the time of instability while white pixels are stable. The left, middle, and right columns represents particles launched from $x=3,5,10$, respectively; in all panels $(y,z)=(0,0)$. The upper row shows cuts in velocity space in the $v_x$-$v_y$ plane, for orbits with $v_z=0$ (purely planar orbits); the lower row shows cuts in the $v_y$-$v_z$ plane, for orbits with $v_x=0$ (initial velocity tangential). Here a blue line indicated the borders of the classical loss cone region.

\begin{figure*}
\begin{center}
\includegraphics[width=\textwidth]{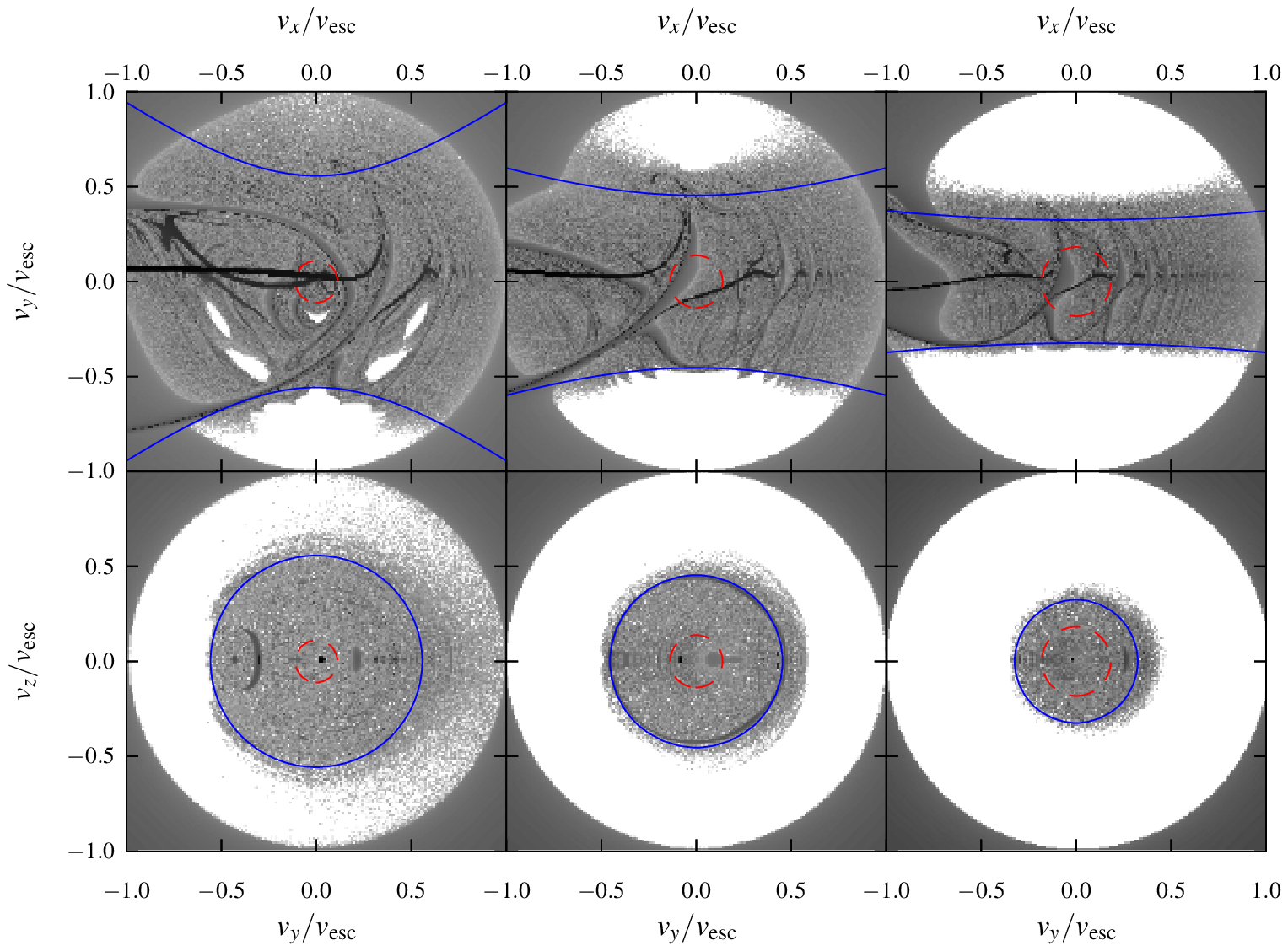}
\includegraphics{./colorbar.eps}
\end{center}
\caption{Same as Fig. \ref{fig:stability-maps-main} for a $q=0.1$ BBH. Axes are normalized by $v_{\rm esc}$ ($=0.96, 0.77, 0.59$ for $x=3, 5, 10$). Note that despite the factor of 10 decrease in the secondary BH mass
compared to the $q=1$ case, the loss cone in fact gets larger (see text), and significant progradre/retrograde asymmetry remains.}
\label{fig:stability-maps-ratio}
\end{figure*}

With decreasing launching radius, the radial orbits get more depleted, and the tangential orbits develop a growing asymmetry. At $x=3$, only retrograde planar orbits remain.

These plots demonstrate that the ``loss cone'' term, coined by \citet{1976MNRAS.176..633F} is inaccurate. The unstable orbits occupy a {\it loss cylinder}, as pointed out by \citet{1978ApJ...226.1087C}. The loss cylinder grows asymmetrically towards the centre, and dominates most of the phase space volume at $r<5$.

The least stable orbits, apart from those on direct collision paths with one of the BHs' tidal radii, are the retrograde orbits just outside the boundary of the classical loss cone, noted by the lower blue lines in Figs. \ref{fig:stability-maps-main} and \ref{fig:stability-maps-ratio}; these orbits appear as dark grey pixels in the stability maps, as they escape on short time-scales. These orbits lead to head on collisions with the approaching BH, and the stars are flung out immediately. However, once the pericentre distance of the retrograde orbit is $r_{\rm peri}>r_{\rm min}$ (orbits below the lower blue line), i.e. the orbit is outside the classical loss cone, the BHs cannot deflect the star appreciably, and the orbit becomes stable on long time-scales. Torquing by the binary on the star changes rapidly due to their large relative velocity with the star, and the net torque tends to cancel out in each close approach of the star.

In contrast, prograde orbits close to the classical loss cone are flung out on longer time-scales. In this case the star approaches the receding BH and the relative velocity is small; the star is subject to a nearly steady torque by the binary, which leads to some energy gain. Repeated close encounters build up the energy of the star, until it is ejected. The further away the pericentre distance is, the smaller is the energy gain, and the longer it takes the star to build up the ejection energy. This leads to a gradual increase in the escape time, moving away from the prograde classical loss cone boundary. Another possible scenario is that prograde orbits close to the classical loss cone do not gradually build up their energy, but rather are on quasi-regular orbits with a nearly fixed energy, i.e. on chaotic orbits with a longer Lyapunov time, which increases with distance from the centre. Once the orbit becomes chaotic, a disruptive head on collision with one of the BHs is quick and inevitable.

The same concept is also illustrated in Figure \ref{fig:stability-maps-ratio}, but there $q=0.1$ and thus $r_{\rm min}=R_2\approx 1.8$, the orbital radius of the secondary BH (see equation \ref{eq:R2}). The larger $r_{\rm min}$ in the $q=0.1$ case, leads to a larger loss cone region, relative to $v_{\rm esc}$, and thus to a smaller area in phase space available for stable orbits. Thus, a $q<1$ binary leads to a larger effect on the orbits stability compared to a $q=1$ binary. However, in the limit $q \rightarrow 0$ the effect of the secondary clearly disappears.

\subsection{The Projected Kinematic Signature}\label{sec:kinematics}

\subsubsection{Monte Carlo Simulations}\label{sec:mc}

Four MC simulations were carried out to calculate the projected LOSVDs. To increase the statistics for the kinematical maps shown below, we took a number of snapshots of the system at different times; all snapshots were taken at times in which the BHs had the same orbital phase (position on the $x$-axis). The snapshots were taken close to the end of the integration at $t=10^4$, where the system was closest to a steady state solution, and were spaced in time by one BBH revolution to ensure significant offsets of the stars between snapshots.

The simulations {\tt main} and {\tt main$\tt {}_2$} are for $q=1$. In {\tt main}, the binary is observed along the $y$-axis (edge-on) and the BHs are maximally separated to the observer (side-view). In {\tt main$\tt {}_2$} the view is along the $x$-axis so the BHs are along the line of sight (front-view). In these two simulations $\sigma =0.25$ (see Table \ref{tab:units}). Simulation {\tt ratio} is for $q=0.1$, for an edge-on side-view. In this calculation $\sigma\approx0.11$. The simulation {\tt single} is for a single BH, located at the origin (i.e. $q=0$); we used $\sigma=0.25$ for it. Note that for this case, the binary separation $a$ has no meaning, and thus there is another free parameter to determine the physical scales of the system.

Table \ref{tab:sims} lists for each simulation the total number of orbits integrated, where $N$ (inner) and $N$ (outer) refer to orbits initiated at $r<2$, and $2<r<60$, respectively (see \S\ref{sec:simulations}). The values of $N$ were set to be large enough to minimize the statistical noise in the results, and may be well above the number of stars contributing to the observed stellar spectral features in typical galaxies.

Table \ref{tab:sims} also gives the fraction of the orbits which are classified as divergent (reaching $r_{\infty}$) or crashing (reaching $r_{\rm tidal}$). This fraction was calculated by giving the weights to the inner and outer datasets, expected from the $r^{-2}$ density profile. The $q=0.1$ simulation has $\sim 30\%$ more diverging orbits, and twice the fraction of crashing orbits than the $q=1$ simulation. However, the total bulge masses of the models are different (depending on $\sigma$), and therefore these fractions translate into different mass deficits. For the $q=1$ simulations, the total bulge mass at $r_{\rm max}=60$ was $M\approx7.42$ (see equation \ref{eq:isothermal-mass}). The relative mass deficit due to diverging orbits in these simulations is therefore $M_{\rm def}/M_{12}\approx 1.01$, where $M_{12}$ is the combined BH mass, $1+q$. For the $q=0.1$ simulation, the total mass at $r_{\rm max}$ was $M \approx 1.35$, and the mass of diverging orbits was $M_{\rm def}/M_{12}\approx 0.45$. The larger fraction of diverging orbits for $q=0.1$ results from the lower bulge contribution within the binary orbit, as expressed by the lower $\sigma/V$, which allowed more orbits to reach $r > r_\infty$. As expected, the two $q=1$ simulations yield the same fraction of divergent and crashing orbits, as the difference is only in perspective, and the statistical error is small enough.

The fraction of $10^{-5}$ orbits which diverged in the $q=0$ simulation results from the numerical error in energy conservation, which allowed this fraction of orbits with $v<v_{\rm esc}$ to become unbound. The fraction of crashing orbits is a factor of 3 to 6 times smaller than in the $q>0$ simulations. This demonstrates qualitatively the enhancement of a BBH on the tidal disruption rate (e.g. \citealt{2008ApJ...676...54C}; \hyperlink{chen09}{\textcolor{Blue}{2009}}), though the exact numbers are not valid given the high value of $r_{\rm tidal}$ used here. In the $q=0.1$ simulation, the primary tidally disrupted 67 times more stars than the secondary.

\begin{table}
\caption{The properties of the four MC simulation carried out. $N$ is the number of orbits integrated in each run (outer and inner regions were calculated separately, see \S\ref{sec:simulations}). The number of snapshots used in the Figures of \S\ref{sec:kinematics} is indicated in the fourth row. The two bottom rows give the fraction of divergent ($r>r_{\infty}$) and crashing ($r<r_{\rm tidal}$) orbits. These numbers can be converted to relative mass deficiencies (see \S\ref{sec:mc}).}
\label{tab:sims}
\begin{tabular}{|l||c|c|c|c|}
\hline 
 & {\tt main} & {\tt main$\tt {}_2$} & {\tt ratio} & {\tt single}\tabularnewline
\hline
\hline
$N$ (outer) & $6.75\times10^{7}$ & $4.26\times10^{7}$ & $8.22\times10^{7}$ & $1.13\times10^{7}$\tabularnewline
\hline 
$N$ (inner) & $5.86\times10^{5}$ & $6.14\times10^{5}$ & $2.49\times10^{5}$ & $1.48\times10^{5}$\tabularnewline
\hline 
$q$ & 1 & 1 & 0.1 & 0\tabularnewline
\hline 
\# Snapshots & 10 & 10 & 10 & 5\tabularnewline
\hline
\hline
Divergent & .26 & .26 & .34 & $\sim10^{-5}$\tabularnewline
\hline
Crashing & .013 & .013 & .026 & .0043\tabularnewline
\hline
\end{tabular}
\end{table}

\subsubsection{The Velocity Distribution Moments}\label{sec:moments}

To characterize the shape of a LOSVD using a small number of parameters, we expand it to a series of the so called Gauss--Hermite (GH) moments; this procedure is consistent with \citet{1993ApJ...407..525V}. We use a function of the form:
\begin{equation}
\mathcal{L}(v) = \frac{\gamma}{\sqrt{2\pi}\sigma}e^{w^2/2} \left[1 + \sum_{n=3}^N h_n H_n(w) \right]
\end{equation}
where $w=(v-\mu)/\sigma$ is the normalized velocity parameter, $H_n(w)$ is the Hermite polynomial of the $n^{\rm th}$ degree. Note that $\sigma$ here refers to the dispersion of this particular line profile. We find the best-fitting $\gamma$, $\mu$, $\sigma$ and $h_n$ ($n\geq 3$) parameters using the least-squares method.

While the GH moments derived above are not standard moments in the statistical sense, as they are derived by least-squares best-fitting to
the data, rather than by projections on the data, 
these commonly used GH terms are useful to describe the deviation of the velocity profile from a Gaussian. For practical reasons we are mostly interested in the first two deviations, as these are often well measured. The corresponding polynomials are:
\begin{eqnarray}
H_3(x) & = & \frac{1}{\sqrt{3}} \left( 2x^3 - 3x \right)\\
H_4(x) & = & \frac{1}{2\sqrt{6}} \left( 4x^4 - 12x^2 + 3 \right).
\end{eqnarray}
The coefficient of $H_3(x)$, $h_3$, is a measure of the profile's lack of symmetry for reflections with respect to its centroid, and $h_4$ is a measure of the even deviation from the Gaussian shape. However, there can be significant contributions to the profiles from higher moment and we therefore also present below plots of the actual LOSVDs along various positions.

\subsubsection{Projected views}

\begin{figure*}
\includegraphics[width=\textwidth]{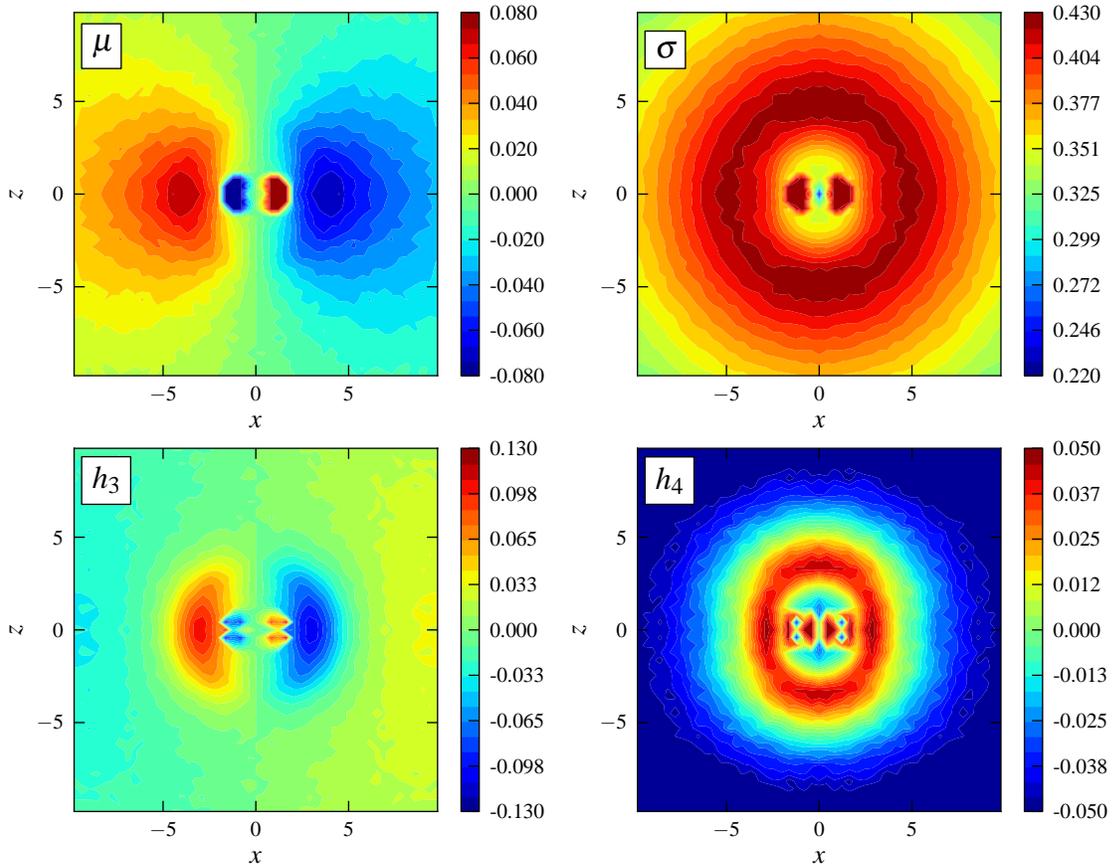}
\caption{Maps of velocity moments of the LOSVD for a $q=1$ BBH. An edge-on side-view of the BBH. The BHs are located at $x=\pm 1, (y,z)=(0,0)$. The line of sight velocity corresponds to $v_y$. Note the ``torus''-like structure in $\mu$ produced by the dominance of retrograde orbits outside the BH Hill spheres. The prograde motion within the Hill spheres reflects the initial conditions. There is a dip in $\sigma$ towards the centre, in contrast to the monotonic rise around a single BH. A torus and a dip are present also in the maps of the higher moments $h_3$ and $h_4$. Note that the peak values of the pixels within the Hill spheres are saturated.}
\label{fig:maps}
\end{figure*}

Figure \ref{fig:maps} presents maps of the projected $\mu$, $\sigma$, $h_3$ and $h_4$ for the $q=1$ simulation; for an edge-on side-view (along the $y$-axis), where the BHs are maximally separated. Each map is composed of $49\times49$ pixels in the projected plane; each pixel represents a moment calculated from the LOSVD measured using 100 bins in velocity. Thus, the spatial resolution is $\approx 0.4$ length units. Due to the maps' up-down symmetry and left-right antisymmetry, and the the fact that each is made of 10 superimposed snapshots, statistics is increased 40-fold. Therefore, these maps are derived from a total effective number of $\approx 3\times 10^9$ particles.

A prominent counter rotating ``torus''-like structure is seen in the top-left $\mu$ map out to a scale 5 times larger than the binary separation. This results from the preferential stability of retrograde orbits, clearly seen in the stability maps (Fig. \ref{fig:stability-maps-main}). On scales of $r\lesssim 2$ one can see the prograde orbits of the stars trapped in the BHs' Hill spheres, as they move together with the BHs.

The top-right panel in Fig. \ref{fig:maps} shows a map of $\sigma$. As expected, $\sigma$ increases inwards, and peaks inside the Hill spheres of the two BHs. However, there is a prominent drop in $\sigma$ at $r \lesssim 3$. This drop occurs at $r$ where all the stable orbits are purely retrograde, and are thus moving relatively coherently around the BBH. Inside the Hill spheres, both prograde and retrograde orbits are allowed, and the $\sigma$ jumps to the expected value for the kinematics around a single BH. The maps of $h_3$ and $h_4$, on the lower left and lower right panels, show similar structures to those seen in the $\mu$ and $\sigma$ maps.

\begin{figure*}
\includegraphics[width=\textwidth]{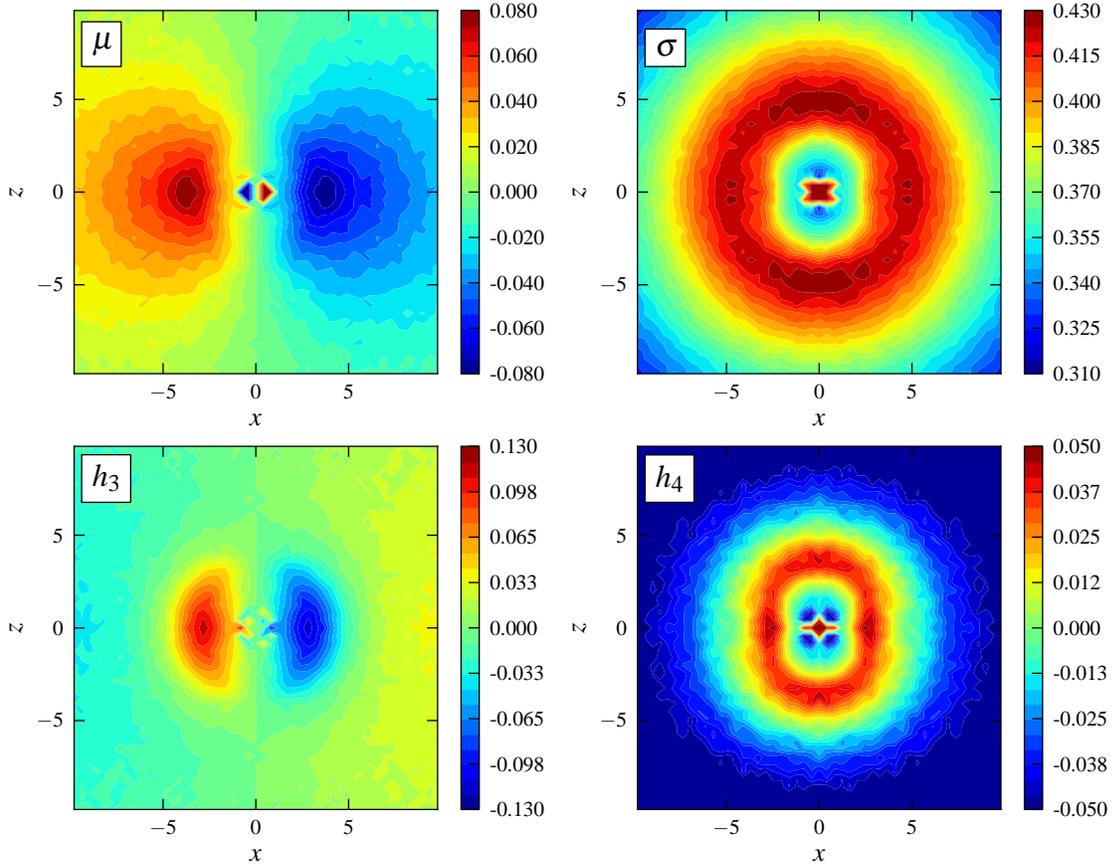}
\caption{The same as Fig. \ref{fig:maps}, for an edge-on front-view, i.e. along the $x$-axis, so both BHs are along the line of sight. As expected, the structure on scales beyond $r\gtrsim2$ remains the same. The Hill spheres now overlap and thus form a more compact structure.}
\label{fig:maps2}
\end{figure*}

Figure \ref{fig:maps2} shows the projected kinematics for an edge-on, front-view (BBH viewed along the binary axis). The maps are remarkably similar to the side-view maps on scales $r>2$. On smaller scales only a single peak is seen in $\sigma$, as the two BHs project on the same position. Interestingly, the $\mu$ map still shows a prograde structure, although the two BHs are moving tangentially to the line of sight. This may result from the non-uniform and anisotropic initial conditions of the bound orbits within the Hill spheres, as measured in
the local corotating frame centred on each BH.

\begin{figure*}
\includegraphics[width=\textwidth]{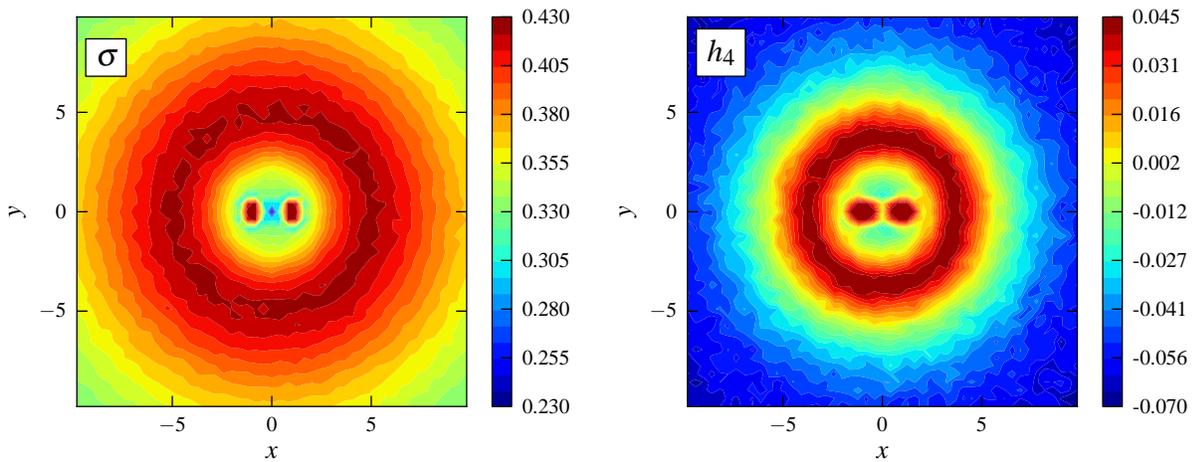}
\caption{The same as Fig. \ref{fig:maps}, for a face-on view (along the $z$-axis). Due to the reflection symmetry of the system with respect to the $x$-$y$ plane, the LOSVD is symmetric, and thus $\mu=h_3=0$. The structure in the maps of $\sigma$ and $h_4$ outside the Hill spheres is similar to that seen in the edge-on views, but now the structure shows a perfect axial symmetry, in contrast to the reflection symmetry in the edge-on views.}
\label{fig:maps-faceon}
\end{figure*}

Figure \ref{fig:maps-faceon} shows a face-on projection of the kinematics (the binary is viewed along the $z$-axis). From symmetry, $\mu=0$ and $h_3=0$ everywhere, and only the maps of $\sigma$ and $h_4$ are shown. Again, a similar structure is seen on scales $r>2$ to that seen in the above two perspectives. The small difference is the perfect axial symmetry, in contrast to the reflection symmetry in the edge-on views.

\begin{figure*}
\includegraphics[width=\textwidth]{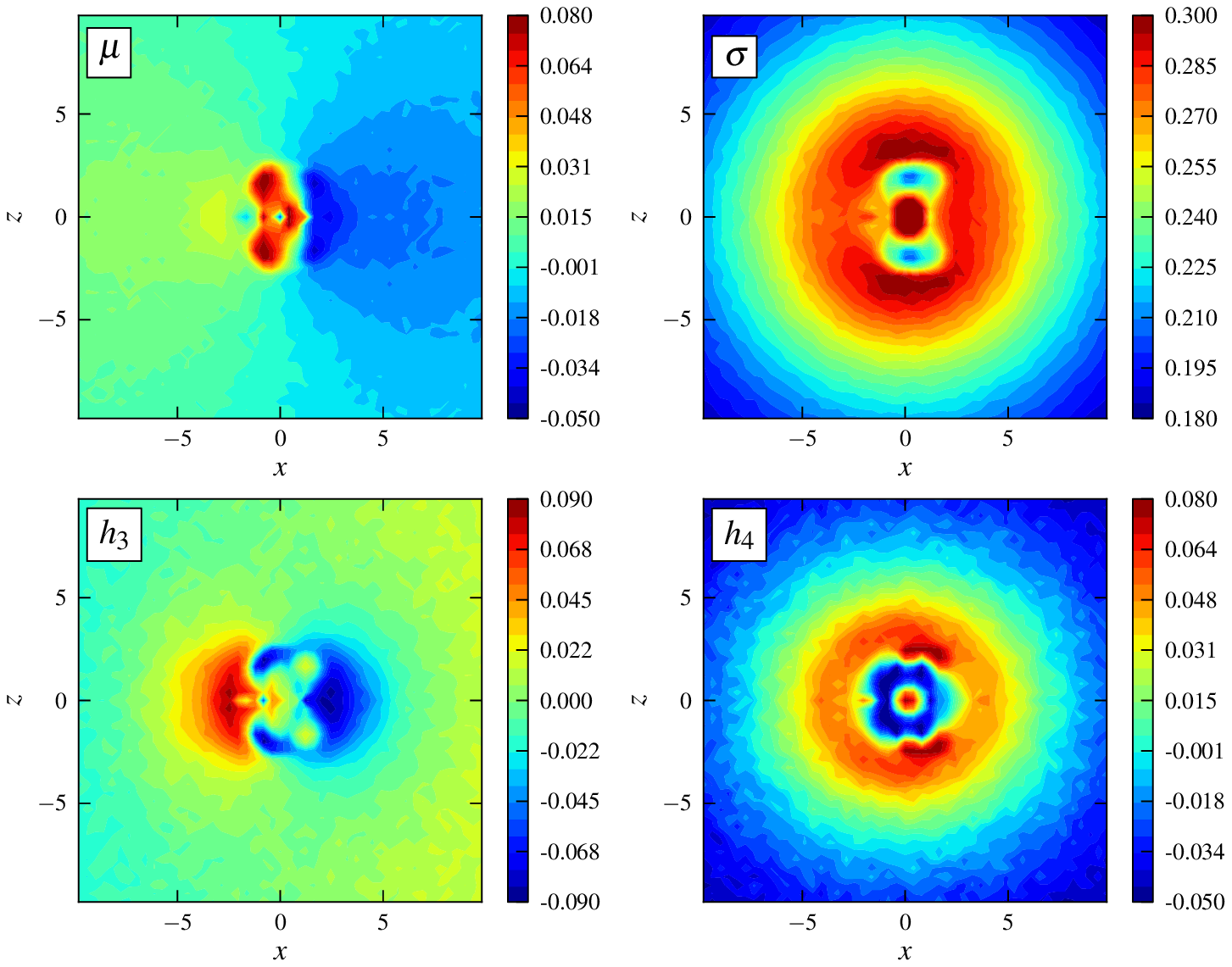}
\caption{An edge-on side-view of a  $q=0.1$ BBH, as in Fig. \ref{fig:maps} for $q=1$. The primary BH is now at $x=0.18$ and the secondary at $x=-1.82$. The structures seen in the $\mu$ and $\sigma$ maps are similar to those seen in the $q=1$ case, but $a$ is now physically 5.5 smaller for the same $M_\bullet$ (Table \ref{tab:units}). The amplitude of $\mu$ in the torus structure, is now lower, in model units, compared to the $q=1$ case, but this is compensated by the higher physical value of the velocity unit. The higher GH moments display more complicated structure, but roughly similar to the $q=1$ case. }
\label{fig:maps-ratio}
\end{figure*}

\begin{figure*}
\includegraphics[width=\textwidth]{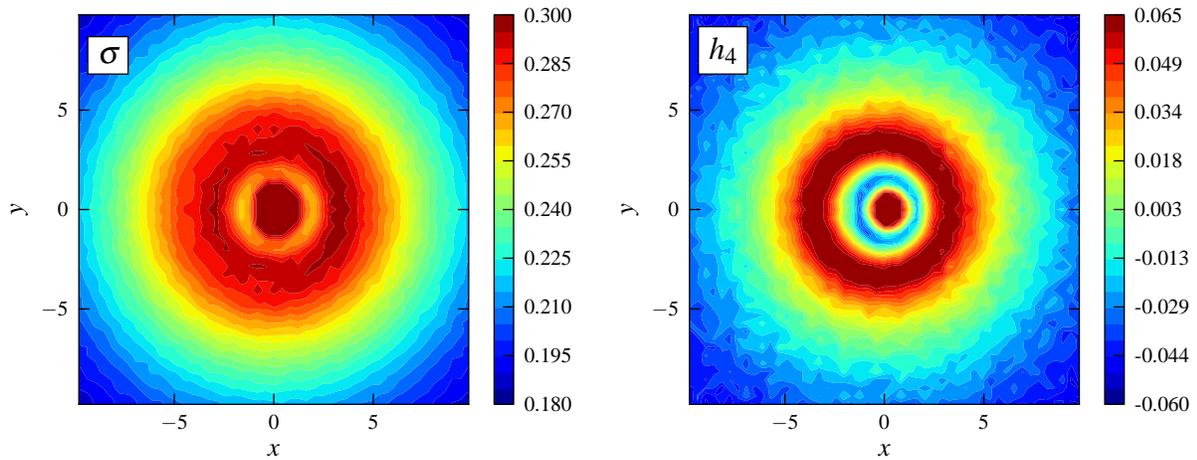}
\caption{A face-on view of a  $q=0.1$ BBH, as in Fig. \ref{fig:maps-faceon} for a $q=1$. The secondary BH at $x=-1.82$ now scours a ``trench'' in $\sigma$ and $h_4$, rather than a wide dip seen in the $q=1$ case.}
\label{fig:maps-ratio-faceon}
\end{figure*}

Figure \ref{fig:maps-ratio} presents an edge-on side-view of kinematics around the $q=0.1$ BBH. The more massive BH is on the right. The kinematic signatures remain prominent, but the structure obviously loses the reflection symmetry of the $q=1$ case. Also, the maps of $h_3$ and $h_4$ become more distinct compared to the maps of $\mu$ and $\sigma$, and not similar as they were in the $q=1$ case. The face-on view in Figure \ref{fig:maps-ratio-faceon} shows that the low mass companion creates a low $\sigma$ ``trench'' along its orbit, and a tiny peak within its tiny Hill sphere. A similar effect is seen in the $h_4$ map.

\begin{figure*}
\includegraphics[width=\textwidth]{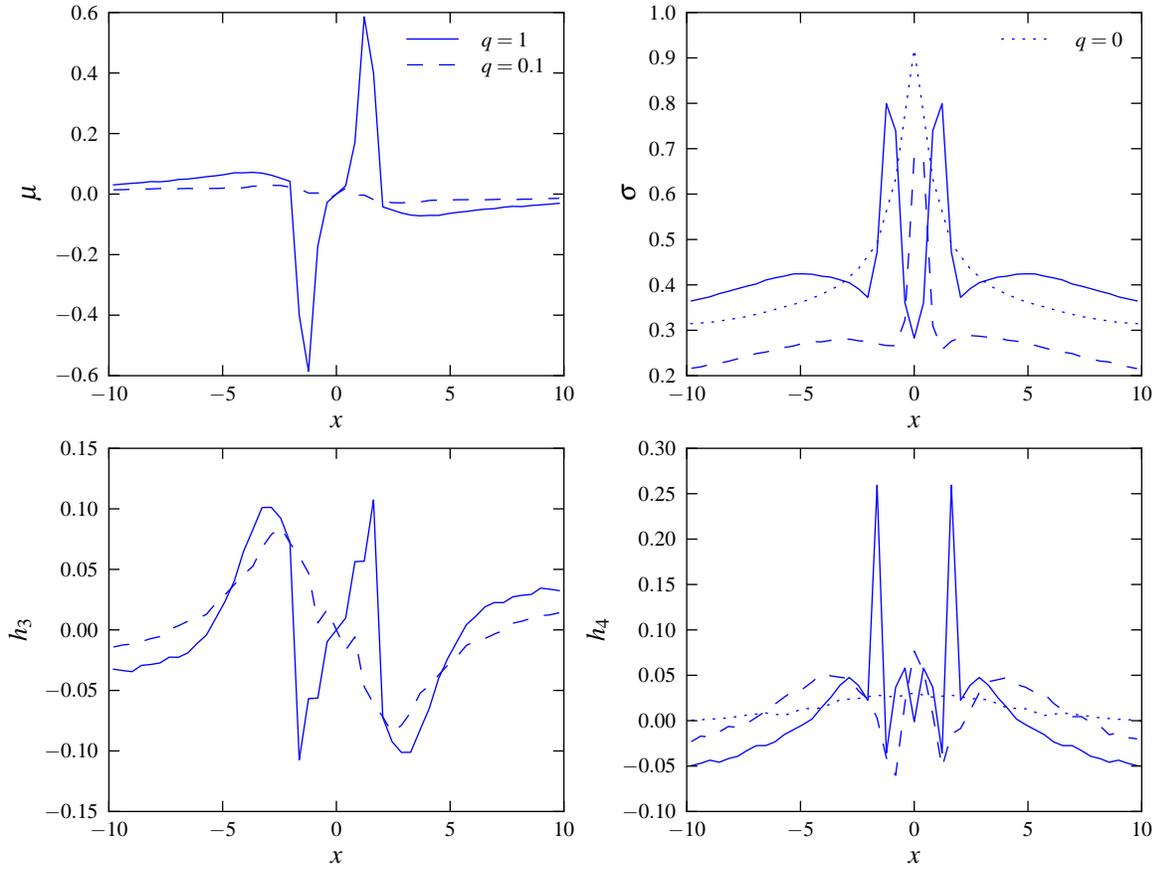}
\caption{The view through a slit of the edge-on side-view maps presented in Figs. \ref{fig:maps}, \ref{fig:maps-ratio}. The slit is placed along the $x$-axis and extends from $z=-0.6$ to $z=0.6$. The $q=0$ result is scaled to $M_\bullet=2$, so a comparison with the $q=1$ can be made. Since $f(v)$ remains isotropic for the $q=0$ case, both $\mu$ and $h_3$ are 0 there. Note the higher $\sigma$ at $\lvert x\lvert>3$ for the $q=1$ case, compared to a single BH of the same total mass. Also, the drop in $\sigma$ at $\lvert x\lvert<5$, in contrast to the $q=0$ case. Large amplitude features are also seen in $h_3$ and $h_4$, in contrast to the $q=0$ case. Note that the horizontal and vertical axes correspond to different physical scales for different values of $q$, for the same $M_\bullet$ (see Table \ref{tab:units}).}
\label{fig:slit}
\end{figure*}

\begin{figure*}
\includegraphics[width=\textwidth]{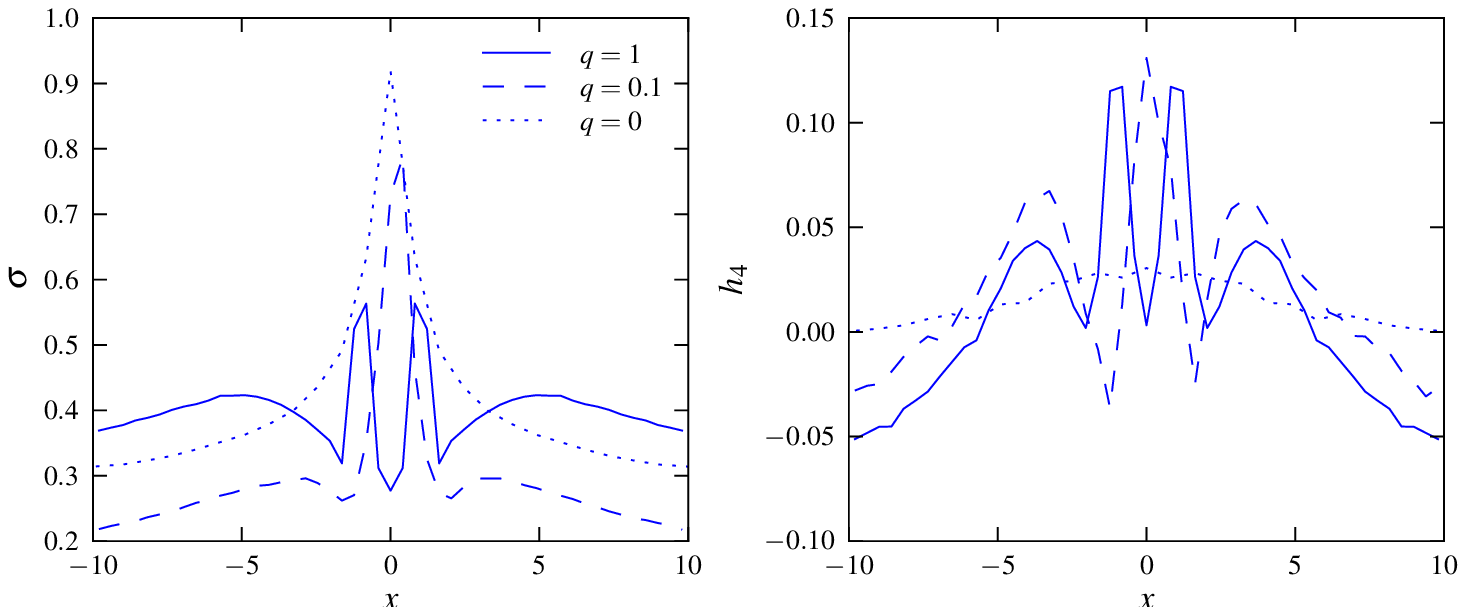}
\caption{ The same as Fig. \ref{fig:slit}, but the system is viewed here along the $z$-axis (``face-on''). The odd moments, $\mu$ and $h_3$, are not shown as they are zero due to symmetry.}
\label{fig:slit-faceon}
\end{figure*}

\subsubsection{Slit Views}

Figure \ref{fig:slit} presents slit views of the four velocity distribution moments. The slit is placed along the $x$-axis; and provides a cross section of the edge-on side-view kinematical maps. The solid lines is for the $q=1$ case, and the dashed line for the $q=0.1$ case. The top left panel present $\mu$. The two sharp peaks at $x=\pm 1$ for $q=1$ represent the cluster of stars trapped in the Hill sphere of each BH, moving at approximately the BHs orbital velocities of $\pm 0.5$. Further out, at $x\approx\pm 3$, there are the broad peaks of the larger scale counter rotating torus, as seen in Fig. \ref{fig:maps}. At a low angular resolution, the broad counter rotating torus structure may be masked by the compact corotating clusters. The amount of dilution depends on the compact clusters luminosity compared to the torus stellar luminosity, which may be different than assumed here.

Interestingly, when $q=0.1$, the two sharp peaks disappear. The massive component, located at $x=R_1\approx 0.2$, now moves at a velocity of only $v_1=q/\sqrt{2(1+q)}\approx 0.067$, below the torus peak velocity. The secondary BH moves faster now, with $v_2=1/\sqrt{2(1+q)}\approx 0.67$, but its Hill sphere volume is now roughly $q^{-3/2}\approx 30$ times smaller, and its contribution to the total profile, given the assumed $\rho(r)$, is negligible.

The top right panel of Fig. \ref{fig:slit} presents $\sigma$ along the slit. The dotted line is for a $q=0$ single BH, with a total mass scaled to 2, which shows the expected monotonic rise as $\sigma\propto r^{-1/2}$ towards the centre. In the $q=1$ case, $\sigma$ at $x>3$ is {\it larger} by 20--40\% than for the single BH case, with the same total mass. This rise occurs because of the exclusions of the low velocities within the loss cone. Furthermore, at $x<5$, $\sigma$ starts falling, in contrast to the sharp rise for the $q=0$ case. This results from the gradual elimination of the prograde orbits with decreasing $r$. This leaves only retrograde orbits, and a quasi-coherent flow, and thus a lower $\sigma$. The double peaks at the centre are due to stars bound within each Hill sphere. For $q=0.1$ only one peak is prominently seen, as expected, since the volume of the secondary Hill sphere is $\sim 30$ times smaller. The radial profile of $\sigma$ also shows a small excess compared to the single BH with the same total mass\footnote{the plotted $q=0$ can be scaled to a total mass of 1.1, for the $q=0.1$ case, by multiplying the velocities by $\sqrt{2/11}\approx 0.43$, and distances by $121/40\approx 3$.}, and a drop close to the centre, but the effects are less pronounced then for the $q=1$ case.

The lower left panel shows $h_3$ along the slit. Comparable peak values are seen in both the $q=1$ and $q=0.1$ cases, indicating that the amplitude of the LOSVD asymmetry is driven by the presence of a second BH, and is not sensitive to its mass for $q=$ in the range 0.1--1. This can also be seen in the stability maps, which show similar asymmetry for $q=1$ and $q=0.1$. The value of $h_4$ along the slit is shown in the lower right panel. The differences from the single BH case are more prominent, and $q=0.1$ presents the largest effects.

Figure \ref{fig:slit-faceon} presents the slit results for a face-on view. The excess in $\sigma$, and its depression at $|x|\lesssim 4$ is clearly apparent for the $q=1$ binary, and a somewhat enhanced depression is also seen for the $q=0.1$ binary, compared with the edge-on view. Significant deviation of the $h_4$ profile from the single BH are now prominent for both $q$ values.

\subsection{Density Profile}

\begin{figure*}
\includegraphics[width=\textwidth]{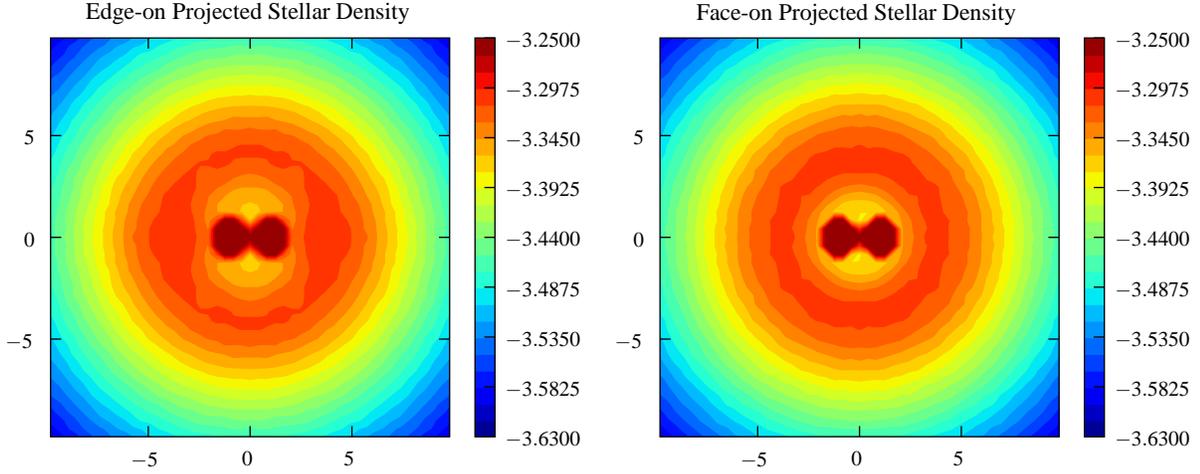}
\caption{The projected density for the $q=1$ simulation. The left panel is an edge-on side-view, and the right panel is a face-on view. Both images are remarkably similar, showing the well known ``scouring'' effect related to the BBH formation. The slight differences is the perfect axial symmetry in the face-on view, and the somewhat elongated core structure in vertical direction in the edge-on view. Colour represents the {\it logarithm} of the {\it normalized} density. The values in the inner regions are saturated.}
\label{fig:projected-density}
\end{figure*}

\begin{figure*}
\includegraphics[width=\textwidth]{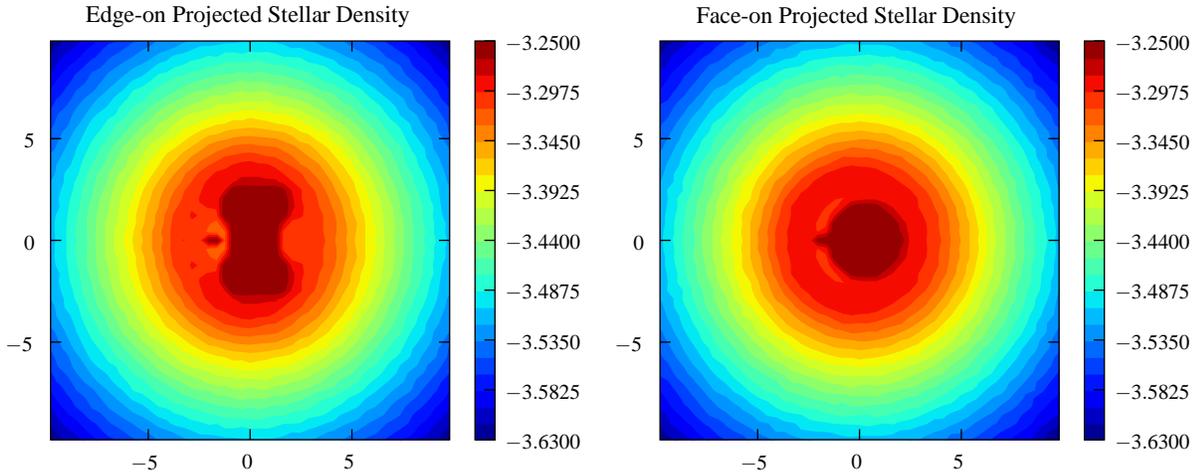}
\caption{The same as Fig. \ref{fig:projected-density}, for a $q=0.1$ BBH. The secondary BH carves out a low density torus structure along its orbit around the primary, producing a circular structure extended in the vertical direction for the innermost stellar light distribution around the primary BH. The face-on view shows an axially symmetric structure with a tiny density enhancement from stars bound to the Hill sphere of the secondary BH.}
\label{fig:projected-density-ratio}
\end{figure*}

Figures \ref{fig:projected-density} and \ref{fig:projected-density-ratio} present an edge-on side-view, and a face-on view of the surface stellar densities for the $q=1$ and $q=0.1$ cases. Both images for the $q=1$ case are remarkably similar, showing the well known ``scouring'' effect of the BBH which depletes stars close to the binary. The slight difference is the perfect axial symmetry of the core structure in the face-on view versus the somewhat elongated core structure in vertical direction in the edge-on view. In the $q=0.1$ case the edge-on side-view shows that the secondary BH carves out a low density torus structure along its orbit around the primary, producing a circular structure extended in the vertical direction, for the innermost stellar light distribution around the primary BH. The face-on view shows an axially symmetric structure with a tiny density enhancement from stars bound to the Hill sphere of the secondary BH.

\begin{figure}
\includegraphics[width=\columnwidth]{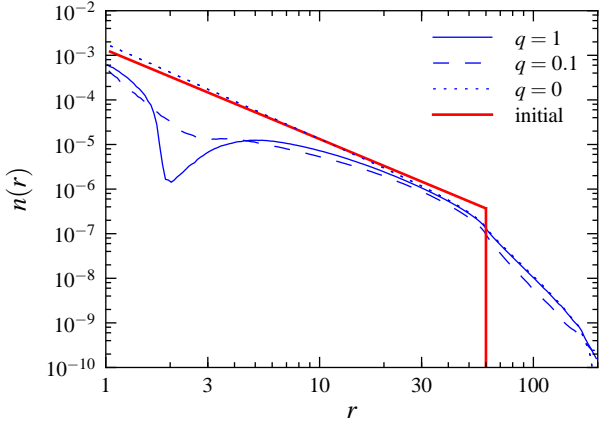}
\caption{The radial stellar density distributions for different $q$ values. $n(r)$ is the normalized number of particles per unit volume. The red line is the initial $\propto r^{-2}$ distribution in the $r=1$--60 range. In the $q=0$ case the distribution remains nearly unchanged, apart from an extension to $r>60$ by stars on nearly radial orbits. The $q=1$ model shows the flattening of $\rho(r)$ at $r<5$, and the deep minimum  at $r=2$, just outside the Hill spheres of both BHs. In the $q=0.1$ case the dip is significantly reduced, partly due to the averaging over spherical shells of a stronger dip confined only close to the secondary BH orbit (see Fig. \ref{fig:projected-density-ratio}).}
\label{fig:density}
\end{figure}

Figure \ref{fig:density} shows the radial density profile $\rho(r)$ averaged along spherical shells. The MC initial condition is $\rho(r)\propto r^{-2}$ out to $r=R_{\rm max}=60$, as noted by the red line in the figure. In the $q=0$ simulation the slope remains close to $-2$ in the inner parts. The slight steepening of the slope towards $R_{\rm max}$, and the much steeper slope beyond $R_{\rm max}$, are edge effects which come from the fact that particles on radial orbits from the initial $r<60$ sphere move out to $60<r<r_\infty$. 

For the BBH cases, the slope flattens inwards towards the core, as the fraction of unstable orbits increases with decreasing $r$. In the $q=1$ case, a dip forms for $r<5$, with a minimum at $r\approx 2$. The rise inward at $r<2$ is produced by stars bound within the Hill spheres of the two BHs. In the $q=0.1$ case the dip is shallower, as the averaging over spherical shells dilutes the density drop which occurs only in the region close to the secondary BH. The difference between the initial and final density distributions is a measure of the stars lost from the system as a function of $r$, as noted in \S\ref{sec:mc}. It is important to note that although our initial condition has a core radius $h=1$, the final state core radius is $\sim 10$, as shown in Fig. \ref{fig:density}, which is closer to the observed core radii in core ellipticals.

\subsection{Internal Kinematics}

\begin{figure}
\includegraphics[width=\columnwidth]{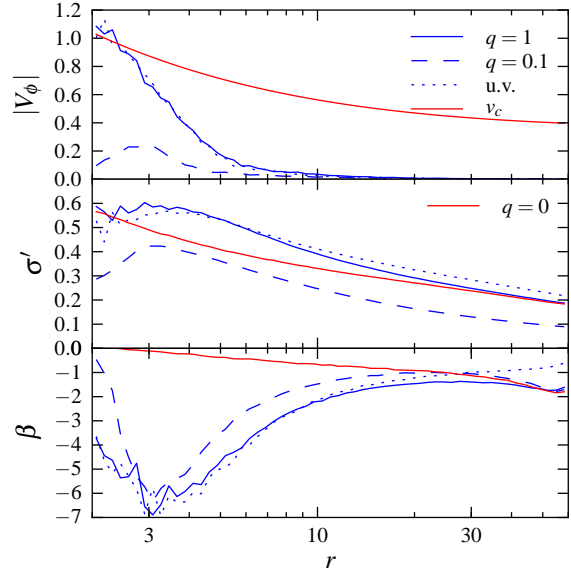}
\caption{The radial dependence of the three internal kinematics parameters. The top panel is the mean tangential velocity. At $r<10$ the preference for retrograde orbits becomes significant, leading to the rise in $V_{\phi}$, reaching a pure circular velocity ($v_c$) at $r<3$, as only retrograde orbits remain. The middle panel presents the 3D rms velocity dispersion. At large $r$ the $q=1$ model matches the $q=0$ case (where $M_{\bullet}=2$), but at smaller $r$ there is a 20--30\% excess (see also Fig. \ref{fig:slit}) produced by the BBH modification of $f(v)$. The bottom panel is the anisotropy parameter $\beta$. The $q=1$ and $q=0.1$ models show the velocities become tangential close to the centre, in sharp contrast to $q=0$, where $f(v)$ remains nearly isotropic at all $r$. The model u.v., shown in all panels, presents a modified $q=1$ model with a uniform velocity distribution extending to $v_{\rm esc}$ (effectively Maxwell--Boltzmann with $\sigma\rightarrow \infty$). The remarkable similarity to the standard $q=1$ model where $\sigma=0.25$ (Table \ref{tab:units}), for all three parameters, shows that the BBH kinematic signature is nearly independent of the form of $f(v)$ used for the initial conditions.}
\label{fig:internal}
\end{figure}

Figure \ref{fig:internal}, upper panel, shows the mean tangential component of the velocity, $V_{\phi}$, as a function of $r$. At $r>10$ both prograde and retrograde orbits are similarly stable, and thus $V_{\phi}$ approaches zero. At $r<10$ retrograde orbits become significantly more stable than prograde (Fig. \ref{fig:stability-maps-main}), leading to a sharp rise in $V_{\phi}$. For $q=1$, $V_{\phi}$ rises from $\sim10\%$ of the circular velocity at $r=8$, to $\sim50\%$ at $r=4$ to $\sim100\%$ at $r=2$, where only retrograde orbits survive. The solid red line is the circular velocity, for comparison; it was calculated assuming a BH mass of 2 in the centre, as in the $q=1$ case. For $q=0.1$, the asymmetry between retro- and prograde orbits remains significant, as can be seen in Fig. \ref{fig:stability-maps-ratio}, but averaging over spherical shells dilutes the effect of the secondary BH, due to the non-axisymmetric morphology of the velocity field in this case, as evident in Fig. \ref{fig:maps-ratio}.

The middle panel of Fig. \ref{fig:internal} shows the 3D velocity dispersion $\sigma'\equiv[(\sigma_r^2+\sigma_\phi^2+\sigma_\theta^2)/3]^{1/2}$ (not to be confused with the 1D velocity dispersion, $\sigma$). The flattening and drop at $r<3$ results, as noted above, from the disappearance of prograde orbits, which leads to a more coherent flow with only retrograde orbits, and thus a lower dispersion. The red line indicates $\sigma'$ for the $q=0$ simulation, for a single BH with a mass of 2. The excess of the BBH 1D $\sigma$, noted in the slit view, can also be seen here for the 3D $\sigma'$.

The bottom panel of Fig. \ref{fig:internal} shows the anisotropy parameter, $\beta\equiv 1-\sigma_t^2/\sigma_r^2$, where $\sigma_t\equiv(\sigma_\phi^2 + \sigma_\theta^2)^{1/2}$ is the tangential velocity dispersion, and $\sigma_r$ is the radial velocity dispersion. The velocity field is significantly anisotropic already at $R_{\rm max}$, as orbits within the loss cone are excluded, leading to $\sigma_t>\sigma_r$ throughout the shown $r$-range. As the loss cone grows inwards, the velocities become more tangential, and thus $\beta$ becomes more negative. At $r<3$ only retrograde orbits remain and the tangential orbits become more coherent, which reduces $\sigma_t$ and increases $\beta$ inwards.

The anisotropy derived here is about an order of magnitude larger in absolute value than in \citet{mm01} (see their figure 16). We suspect that this issue results from the fact that the BBH in that work does not stall at $a_{\rm h}$ but rather still shrinks rapidly at this radius (as indicated by figure 1 there). The stellar distribution there cannot reach a near steady state solution, as derived here. We verified that a shorter integration time in our simulation indeed yields a $\beta$ which is smaller by factor of few. Thus, the stellar kinematics derived in \citet{mm01} does not capture the full level of the kinematic signature of a BBH which is stalled at $a_{\rm h}$.

The dotted line model, shown in all panels, presents a modified $q=1$ model with a uniform velocity distribution extending to $v_{\rm esc}$, or effectively a Maxwell--Boltzmann distribution with $\sigma\rightarrow\infty$. The remarkable similarity to the standard $q=1$ model, where $\sigma= 0.25$ (Table \ref{tab:units}), for the three internal kinematics parameters, shows that the BBH kinematic signature is nearly independent of the form of $f(v)$ used for the initial conditions.

\subsection{LOSVDs}

\begin{figure*}
\includegraphics[width=\textwidth]{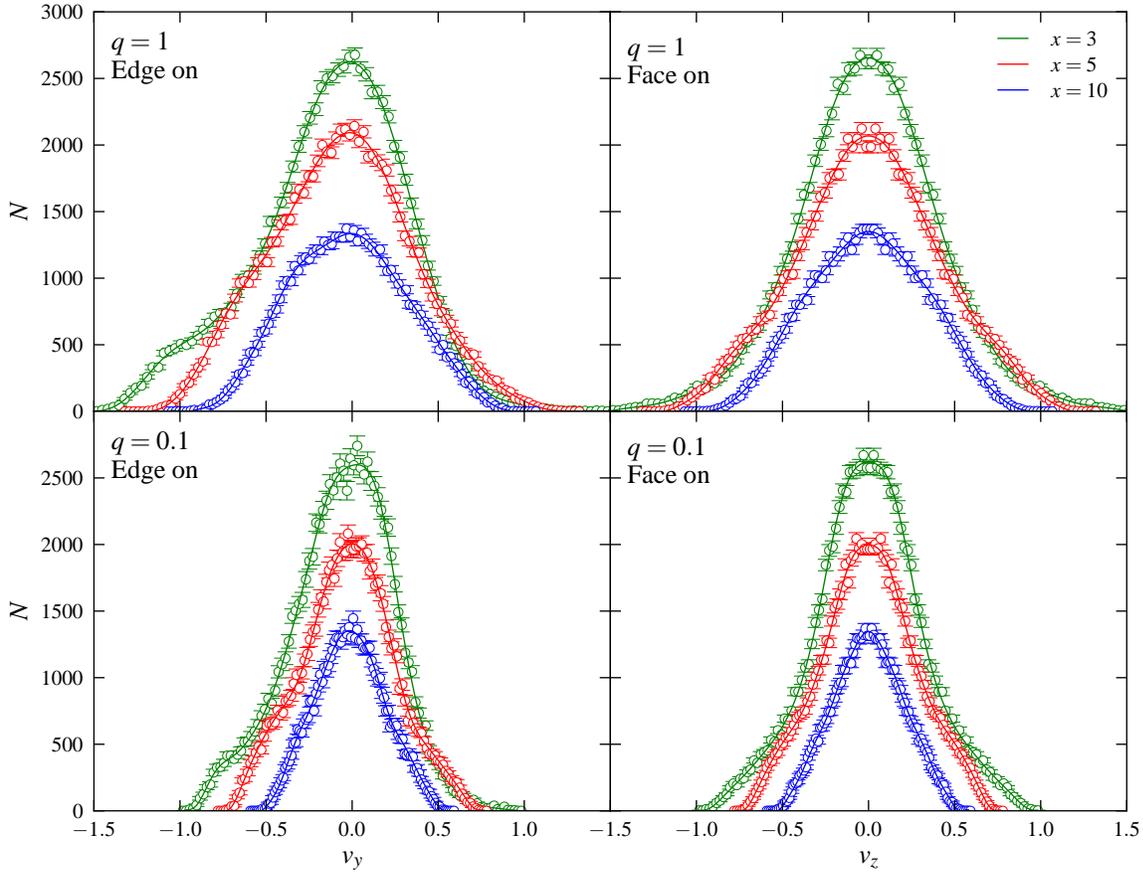}
\caption{The LOSVDs at different positions. The viewing angle and the model are noted in each panel. Each panel shows three lines of sight (along the $y$-axis) at $z=0$ and $x=3,5,10$. The error bars are statistical errors. The solid lines are the GH best-fits up to order 20. Note that the line FWHM changes by $\sim 10$\% moving inward from $x=10$ to $x=5$, for the $q=1$ case, and {\em drops} moving inwards to $x=3$, in sharp contrast to expected rise for the single BH case. In the $q=0.1$ case the FWHM increases from $x=10$ to $x=5$, but remains constant from $x=5$ to $x=3$, in contrast to the expected rise for a single BH. A noticeable asymmetry near the line base is seen for both $q$ values. The lines are symmetric in the face-on views, and show similar trends in the FWHM vs. distance from the centre, as seen in the edge-on, side-view. }
\label{fig:losvd}
\end{figure*}

Figure \ref{fig:losvd} shows the LOSVD along certain lines of sight. Each panel shows the LOSVD along three lines of sight at distances of 3, 5, and 10 from the centre of gravity (origin of the coordinates). The size of the aperture corresponds to a projection pixel, an area of $0.4\times 0.4$ square length units. The top panels are for the $q=1$ case, and the bottom panels are for $q=0.1$. The left panels are for an edge-on side-view of the binary and the right panels are for a face-on view. The vertical axis is in units of orbits per velocity bin, where each line has 100 velocity bins; the range of velocities is determined according to the maximal escape velocity along each line of sight, the same values used in the stability maps (Figs. \ref{fig:stability-maps-main} and \ref{fig:stability-maps-ratio}).

The line FWHM changes by $\sim 10$\% moving inward from $x=10$ to $x=5$, for the $q=1$ model, and it drops moving further inwards to $x=3$, as also seen in the slit views of $\sigma$ (Figs. \ref{fig:slit}, \ref{fig:slit-faceon}). This is in sharp contrast to the single BH case, where the FWHM is expected to rise by 83\% ($=\sqrt{10/3}$) from $x=10$ to $x=3$. Note also the line asymmetry in the edge-on view, which increases moving inwards, reflecting the enhanced retrograde motion close to the centre. In the $q=0.1$ case the presented lines of sight are away from the secondary BH, at $x=-1.82$, somewhat reducing the profile asymmetry. The FWHM of the lines increases from $x=10$ to $x=5$, as expected for a single BH, but it remains constant from $x=5$ to $x=3$ (as seen in Figs. \ref{fig:slit}, \ref{fig:slit-faceon}), in contrast to the expected 30\% rise for a single BH. A noticeable asymmetry near the line base is seen for the edge-on view, similar to the asymmetry seen in the $q=1$ case, but with a lower amplitude.

The LOSVDs for $v_z$, seen for the face-on view, are symmetric, as expected due to the reflection symmetry of the system with respect to the $x$-$y$ plane. In contrast, the edge-on $v_y$ profiles are asymmetric due to the prograde/retrograde asymmetry discussed above. The asymmetry increases for lines of sights closer to the centre, and towards the wings in each profile, as these are produced by orbits closer to the binary, where the tangential velocity asymmetry becomes larger.

\begin{figure*}
\includegraphics[width=\textwidth]{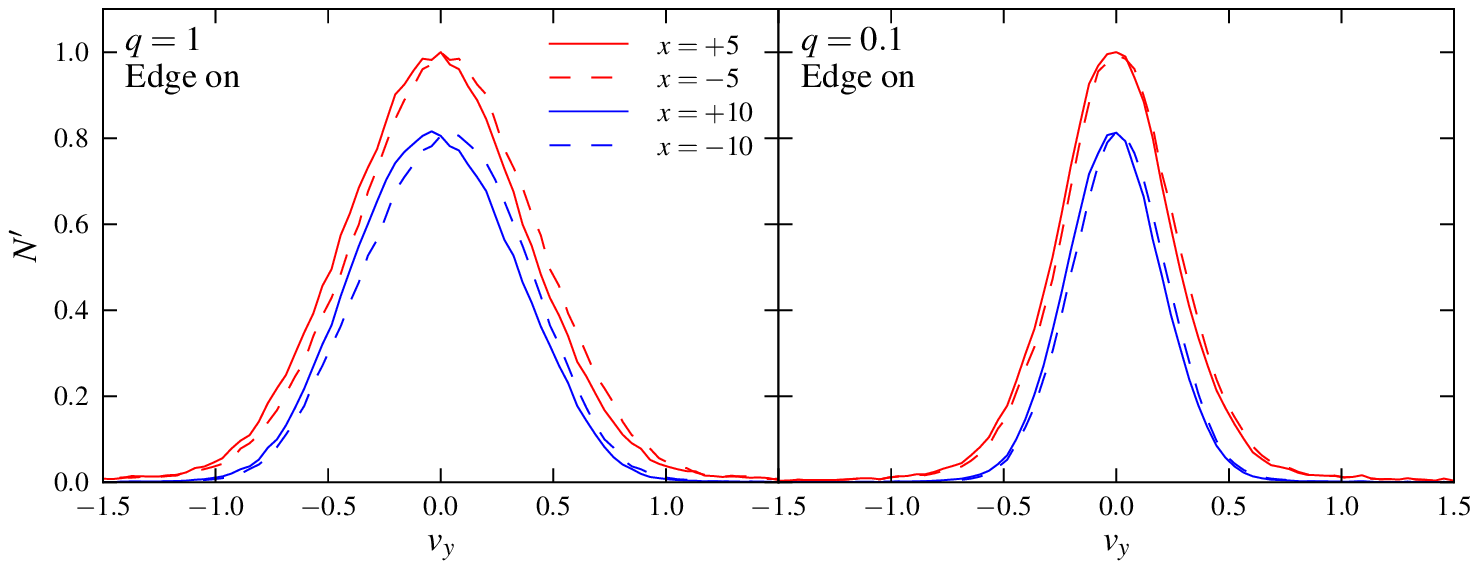}
\caption{The LOSVDs for an edge-on view with a low angular resolution observation. The LOSVDs are integrated through a circular aperture with a Gaussian weight function with a FWHM of 10, representing an angular resolution five times worse than the binary separation. The left panel is for $q=1$ and the right for $q=0.1$. The lines of site are centred at $x=\pm 10$ and $x=\pm 5$, i.e. on both sides of the BBH. There is a clear shift in the line peaks in the $q=1$ case between the two sides, resulting from the rotation structure evident in the side-view maps (see Fig. \ref{fig:maps}). The shift is significantly smaller in the $q=0.1$ case.}
\label{fig:losvd-wide}
\end{figure*}

Figure \ref{fig:losvd-wide} shows the LOSVDs expected from a low angular resolution observation for an edge-on line of sight. We compare the velocity profiles for two lines of sight situated on opposite sides of the centre, at distances of 5 and 10 from the centre. We integrated the LOSVD through a circular aperture using a Gaussian with a FWHM of 10 as the weight function, which represents the angular point spread function (PSF) of the telescope. The left panel is for $q=1$ and the right for $q=0.1$. The $q=1$ case shows a clear shift in the peaks of the LOSVDs, which results from the net rotation of the stars, clearly seen in the spatially resolved maps of the average velocity $\mu$ (Figs. \ref{fig:maps}, \ref{fig:maps-ratio}). The shift is more prominent for the lines of sights centred at $x=\pm 10$, i.e. at a position a FWHM of the PSF away from the centre. In the $q=0.1$ case the PSF used here eliminates almost completely the profile differences between the two lines of sight, and the binary kinematic signature is very small.

\section{Discussion}\label{sec:discussion}
The extensive scattering experiments presented above, using $\sim 10^8$ test particles surrounding a massive BBH embedded in a bulge potential, allowed us to accurately map the 3D velocity distribution of stable orbits. These are used to derive maps of the projected 2D velocity distribution moments, and of the LOSVD along various directions. The stable orbits close to the binary are generally tangential and preferentially retrograde, leading to retrograde ``torus'' structure in the projected average velocity. The velocity dispersion shows an excess of 20--40\% compared to a single BH of the same total mass, and shows a dip close to the binary. These effects lead to a clear kinematic signature of the BBH,  which can be detected on scales of 5--10$a$. Thus, they can be spatially resolved even when the binary cannot be resolved, as expected even in the nearest galaxies \citep{yu02}.

Interestingly, the maps of the 2D velocity distribution moments for both the $q=1$ and the $q=0.1$ cases, show that the kinematic signature extends on similar scales (in units of $r/a$) and with similar amplitudes (in absolute velocity). One could expect a smaller effect if the companion BH is of smaller mass, but apparently the stability of orbits is strongly influenced by a secondary with just 10\% of the primary's mass. However, the stalling radius is a factor of $(1+q)/2q=5.5$ smaller in the latter case, and will thus be harder to spatially resolve.

The tendency for counter rotating orbits and a velocity dispersion drop close to the centre were briefly noted by \citet{mm01}. However, their results were based on $N$-body simulations of $\sim 10^5$ particles, carried out on scales $\sim 100$ larger than here, required to simulate the merger of the two bulges. As a result, there were only $\sim 10^3$ particles in their study on the $r_{\rm infl}$ scale (see their table 2). The implied large statistical errors in that study, therefore did not allow to probe the stellar kinematics on the scale of $r_{\rm infl}$ and closer to the BBH probed here. The maps of the projected kinematics produced by \citet{mm01} therefore do not show the BBH signature presented here.

We also find a clear drop in the projected stellar surface density, as stars are efficiently ejected from regions just outside the Hill spheres of both BHs (for $q=1$), or of the secondary BH (for $q=0.1$). This is a well known effect (e.g. \citealt{2001A&A...377...23Z}), and therefore we do not discuss it further here.

The advantage of an $N$-body simulation is that it allows to follow the system, starting with plausible initial conditions of separate bulges, and derive the resulting stellar velocity distributions on large scales following the merger. For example, \citet{mm01} found that stars in the merged bulge have a net rotation in parallel to the initial angular momentum of the two bulges. In the scattering experiments, we just assumed some initial conditions for the stellar distribution, and do not know whether these were plausible. However, since our simulations are much faster than $N$-body, we can explore the dependence of the results on the initial conditions. The initial stellar $f(v)$ used in the MC simulation is drawn from the bulge $\sigma$, and as shown in Figs. \ref{fig:stability-maps-main}, and in particular \ref{fig:stability-maps-ratio}, $\sigma$ can become a small fraction of $v_{\rm esc}$ at small $r$. To explore the dependence of the BBH signature on the initial $\sigma$ chosen for $f(v)$, we repeated the analysis with the extreme assumption $\sigma\rightarrow \infty$, i.e. a uniform $f(v)$ extending to $v_{\rm esc}$. As shown in Fig. \ref{fig:internal}, this uniform initial $f(v)$ produced nearly identical results. The uniform $f(v)$ also led to small deviations ($<10$\%) in the slit view results, indicating that the BBH signature is independent of the form of the initial $f(v)$. Thus, the merger history of the binary, which may set $f(v)$ on larger scales of the bulge, is likely not important on the scales of $r/a\sim$ a few, when the binary becomes hard and stalls.

We also explored the effects of the uniform density core radius $h$ (equation \ref{eq:isothermal-density}), and produced stability maps for $h=3$ instead of $h=1$, for the same $\sigma$. The maps look nearly indistinguishable from those shown in \S\ref{sec:stability-maps}. This is expected as increasing $h$ reduces the bulge mass, which is already very small for $h=1$ close to the BBH (equation \ref{eq:isothermal-mass}). So, the BBH signature is independent of the exact form of the inner bulge potential, as expected since a hard binary resides at $a_{\rm h}\ll {GM_\bullet}/{4\sigma^2}$, where the bulge mass is negligible. However, in a flat core the integrated line of sight stellar light increases outwards, increasing the dilution of the BBH kinematic signature by the extended stellar light. The final state core radius derived here of $H\sim 10$ (Fig. \ref{fig:density}) is significantly larger than the initial condition of $h=1$, but is clearly too small in some core ellipticals, where the BBH kinematic signature will be harder to detect.

Furthermore, as shown in Fig. \ref{fig:stability-maps-longterm}, a large fraction of the unstable orbits are lost on a time-scale of $t<10^3$, or $<80$ BBH periods, which corresponds to a few orbital times for orbits starting at $r\sim$ 5--10. Thus, the BBH kinematic signature is largely imprinted already on a time-scale $<10^6$~yr. Unstable orbits present in the initial $f(v)$ are quickly excluded. The uniqueness of the BBH signature depends on the population of the velocity phase space of stable orbits. Since the stable orbits around a BBH are quasi-periodic, a given orbit likely moves around in velocity phase space, and so a population of stars which populate only a small peculiar corner of the stable region in velocity phase space appears unlikely. The BBH signature is therefore likely well defined.

The unstable orbits are essentially orbits within the loss cone (or ``loss cylinder'', as mentioned above), which acquires a more complicated shape when $r/a$ is a few. The kinematics we described may be more accurately termed ``kinematic signature of loss cone depletion''. The loss cone refilling mechanisms will tend to erase the signature presented above. These mechanisms tend to become more effective on larger $r$, in particular at $r\gg r_{\rm infl}$, where either steady state, or time dependent perturbations to the spherically symmetric potential assumed here are more likely to be found, or to occur transiently. On the scale of $r\lesssim 10$ probed here, such mechanisms are less likely to occur. If they occur transiently, they are less likely to have a significant effect, given the shorter survival time of unstable orbits on these scales. 

The enhancement of the observed $\sigma$ within the BBH $r_{\rm infl}$ by 20--40\% implies that the standard direct estimate of $M_{\bullet}$, which assumes an isotropic $f(v)$, will lead to an overestimate of $M_{\bullet}$ by a factor of 1.5--2. 

Once the BHs coalesced, the signature will be lost on a time-scale of $M_*/\dot{M}_*$, where $\dot{M}_*$ is the loss cone refilling rate and $M_*$ is the stellar mass at $r\lesssim 10$. This will likely occur on a time-scale significantly longer than the dynamical time-scale at $r_{\rm infl}$, and may in fact be longer than the Hubble time, if it occurs in the low density core of a giant elliptical galaxy. However, if the BBH acquires a significant kick following the merger, it may oscillate around the core \citep{2008ApJ...678..780G}, which will erase the BBH kinematic signature on a much shorter time-scale, possibly while enhancing the core due to heating of the stars. The detection of the predicted BBH kinematic signature implies the presence of a BBH currently, or a merger which took place on time-scales shorter than the loss cone refilling time, although this timescale may be the Hubble time in giant ellipticals.

The calculations presented above follow test particles, and thus do not take into account the energy and angular momentum lost from the BBH due to the stellar ejection, which can be significant given the large fraction of ejected stellar mass (see \S\ref{sec:mc}). The justification is that the purpose of this work is to look for the steady state solution, i.e. find which orbits may be populated and which cannot survive for long, when the BBH is at the stalling radius, rather than follow the time evolution of the system. The implied significant energy loss of the BBH found here, results from the inappropriate initial conditions of a spherically symmetric $\rho\propto r^{-2}$ assumed here for a hard BBH. In reality, stars may be ejected from the system much earlier when the binary just becomes bound, i.e. when $a\sim r_{\rm infl}$, or potentially even earlier and on larger scales, based on the high value of $M_{\rm def}/M_\bullet\sim 10$, and the small scatter, found by \citet{2009ApJ...691L.142K} for massive ellipticals.

Here we find that the kinematic signature of the BBH is imprinted on the same scale ($r<10$) that the surface density signature of the BBH is imprinted (inevitable as stars with specific kinematics are lost). Therefore, if $M_{\rm def}$ is indeed a signature of the BBH formation process, then the BBH kinematic signature may be imprinted already on the significantly larger scales of the core radius, where $M_{\rm def}$ is measured, of the order of tens to hundreds of parsecs (\citealt{1997AJ....114.1771F}; \citealt{2009ApJS..182..216K}), and may be more easily detectable, possibly already in existing data. Clearly, it is interesting to explore the BBH merger starting from $r_{\rm infl}$, and follow the resulting kinematic signature on larger scales than those calculated in this study.

The calculations presented here assume circular BH orbits. The BBH may have a high eccentricity due to various mechanism (e.g. \citealt{1993PASJ...45..303M}; \citealt{2007Sci...316.1874M}; \citealt{2007ApJ...660..546S}; \citealt{2009ApJ...695..455B}). In that case the effects calculated here will likely extend to larger scales, set by the major axis of the binary orbit. It is less clear if the binary stalls in such a case, and at what radius, if it does.

It is also interesting to note that stars bound within the Hill spheres preserve the original populations before the merger occurred, as stars outside the Hill sphere with a total energy below the Hill sphere potential barrier (measured in the corotating frame, where the energy of each orbit is conserved) cannot enter it, and stars with a total energy above the potential barrier in the Hill sphere, are on highly unstable orbits and disappear quickly from the system. Stars can enter or leave the Hill spheres only through an energy exchange with a fourth body, which may be very slow in low density cores.

\section{Conclusions}\label{sec:conclusions}
Orbits in the restricted three body problem are notoriously complex (though some insight can be gained from stability maps). Here we exploit this property to derive the signature of a BBH on the nearby stellar kinematics, once unstable orbits are gone. The fraction of velocity phase space populated by stable orbits decreases inwards. The stars ejected by unstable orbits will leave behind a light deficiency, which was suggested to explain the core structure of massive ellipticals. The remaining stars are on significantly anisotropic orbits, characterized by the following properties:
\begin{enumerate}
 \item Tangential orbits dominate, mostly retrograde at $r<5$.
 \item Increased $\sigma$, due to the elimination of low tangential velocity orbits.
 \item A drop in $\sigma$ at $r<5$, as most orbits become retrograde.
\end{enumerate}
These properties lead to a specific signature on the LOSVD moments on scales as large as 5--10$a_{\rm h}$, which may be resolved in nearby galaxies. The detection of these kinematic features may indicate the presence of a BBH currently, or a relaxation time ago, beyond which the kinematic signature is erased. If the core structure is a signature of a BBH phase in the past, some BBH kinematic signature may remain on the core scale as well.

\section*{Acknowledgements}
We thank the referee for helpful comments. We also thank David Merritt, Scott Tremaine and John Kormendy for helpful comments and discussions.

\label{lastpage}

\end{document}